\documentstyle[graphicx,11pt,aaspp4]{article}
\def\Msun{~M_\odot}
\def\lsim{\raise0.3ex\hbox{$<$}\kern-0.75em{\lower0.65ex\hbox{$\sim$}}}
\def\gsim{\raise0.3ex\hbox{$>$}\kern-0.75em{\lower0.65ex\hbox{$\sim$}}}

\def\kms{\rm ~km~s^{-1}}

\def\ml{~\Msun ~\rm yr^{-1}}

\begin{document}

\title{WIND INTERACTION MODELS FOR GAMMA-RAY BURST AFTERGLOWS:
THE CASE FOR TWO TYPES OF PROGENITORS}
\author{Roger A. Chevalier and Zhi-Yun Li}
\affil{Department of Astronomy, University of Virginia, P.O. Box 3818}
\affil{Charlottesville, VA 22903; rac5x@virginia.edu, zl4h@virginia.edu}


\begin{abstract}

Beginning with the $\gamma$-ray bursts GRB 970228 and GRB 970508, a
standard model for the interpretation of GRB afterglows emerged involving
synchrotron emission from a constant energy blast wave expanding into
a constant density, ``interstellar'' medium.
However, a massive star origin for GRBs implies a
stellar wind environment, probably a Wolf-Rayet star, and 
we have previously suggested
wind interaction 
models for the afterglows of GRBs 980326, 980519, and 980425/SN 1998bw.
Here, we extend the theory of afterglows in winds, considering
strong cooling phases, the transition to nonrelativistic motion, jets,
and prompt, reverse shock emission.
We find that, compared to the interstellar case, the optical prompt 
emission in the wind case could have a comparable magnitude, but would 
die off faster.  
We examine the afterglows of other well-observed GRBs in the
context of wind interaction models, and find that GRBs 970228 and 970508
 are likely wind interactors.
The revision in the nonthermal afterglow emission from GRB 970228
caused by the recognition of late supernova emission favors
wind interaction.
The radio evolution of GRB 970508 provides especially strong evidence
for wind interaction.
For GRB 970508, the observations suggest a density that is compatible
with that expected in a Wolf-Rayet star wind.
Finally, observations of the afterglow evolution of 
GRBs  990123 and 990510 and the prompt optical emission from
GRB 990123 favor interstellar
interaction models, which would suggest compact star merger progenitors
for these objects.

\end{abstract}


\section{INTRODUCTION}

The fireball model for GRBs (gamma-ray bursts) led to predictions
of the afterglow emission that might be expected when the energetic
shock wave encountered the surrounding medium (Katz 1994;
M\'esz\'aros, \& Rees 1997).
The subsequent optical and X-ray observations of the afterglow from GRB 970228
appeared to confirm the predictions of the simplest afterglow model
(Waxman 1997a; Wijers, Rees, \& M\'esz\'aros 1997).
This model involved synchrotron emission from electrons accelerated to
a power law energy spectrum in a relativistic blast wave expanding
into a constant density, presumably interstellar medium (ISM).
In particular, the expected relation between the flux spectral index
and the power law rate of flux decay was in approximate accord with
the observations.
This model has become the ``standard model'' for the interpretation
of GRB afterglow observations.
It was used to make predictions of bright optical emission in the early
phases when a reverse shock front is present (Sari \& Piran 1999b).
The observation of a bright flash from GRB 990123 (Akerlof et al. 1999)
gave basic confirmation of this aspect of the model (Sari \& Piran 1999a;
M\'esz\'aros \& Rees 1999).
The expectation of jets for the initial energy deposition led to
predictions of the effects on the light curve as the jet slowed
(Rhoads 1997, 1999; Sari, Piran, \& Halpern 1999).
The observations of the afterglow of GRB 990510 confirmed the
basic features expected for jet deceleration (Harrison et al. 1999).

These successes of the standard model give confidence that it
is essentially correct.
However, there has been increasing evidence that at least
some GRBs have massive star progenitors, as initially 
suggested by Woosley (1993)
on theoretical grounds.
Paczy\'nski (1998) noted that the available evidence on the location of GRBs
in their host galaxies indicated a link to star formation.
A more direct link to massive stars is provided by the presence of a 
supernova, the light
from the exploded star matter.
GRB 980425 was probably associated with the relatively nearby Type Ic supernova
SN 1998bw (Galama et al. 1998d; Kulkarni et al. 
1998), and evidence for supernova type emission has
now been found in GRB 980326 (Bloom et al. 1999b) and GRB 970228
(Reichart 1999; Galama et al. 1999b).

The importance of a massive star origin for the afterglow evolution
is that the GRB blast wave should be expanding into the stellar wind
of the progenitor star.
Dai \& Lu (1998), M\'esz\'aros, Rees, \& Wijers (1998) and
Panaitescu, M\'esz\'aros, \& Rees (1998) described some features of
afterglow evolution in a $\rho\propto r^{-2}$ stellar wind.
Chevalier \& Li (1999, hereafter CL)
made specific estimates for expansion into the wind of a Wolf-Rayet star.
Li \& Chevalier (1999) and
CL identified GRB 980425/SN 1998bw, GRB 980326, and GRB 980519 as likely
circumstellar wind interactors based on their afterglow evolution.
In the cases of GRB 980425 and 980326, this association is supported by
the presence of supernova emission.
In GRB 980519, a supernova would have to be somewhat
fainter than SN 1998bw (Bloom 1999).
CL identified GRB 990123 as a likely interstellar interactor based on its
afterglow evolution.
GRBs of this type are not expected to be accompanied by supernovae.

CL placed the afterglows of GRB 980519 and GRB 980326 in the wind category
partly based on their relatively rapid rates of decline of optical emission.
An alternative explanation for the decline is that the emission is from
a laterally expanding jet (Sari, Piran, \& Halpern 1999; Halpern et al. 1999).
However, in the case of GRB 980519, CL were able to use radio observations
made during the first 3 days (Frail et al. 1998b) to 
further constrain the model and to predict the radio evolution.
Radio data on GRB 980519 over the first 63 days are now available (Frail
et al. 1999) and they are in  agreement with the prediction of CL.
The jet model does not appear to fit the data as well, but scintillation
effects cause sufficiently large uncertainties in the radio fluxes
that the jet model is also acceptable.

In view of the increasing evidence for massive star progenitors of GRBs,
our aim here is to further develop the theory of wind interaction and
to examine the data on afterglows in the context of wind and interstellar
interaction models.
In \S~2, we extend the theory of afterglows in winds to consider
strong cooling, nonrelativistic evolution, and jets.
Our discussion is guided by previous discussions of the constant
density interaction case, and we contrast the two situations.
The differences are expected to be especially significant for prompt emission
because of the large difference in the ambient density at early times.
We treat prompt, reverse shock emission for the wind case in \S~3.
In \S~4, we examine the data on the best observed afterglows in the
context of the wind and constant density models.
We find that wind models can explain a number of observations previously
regarded as puzzling, pointing to wind interaction models for these cases.
Although wind models are indicated for some observed
afterglows, there are others that are better described by
constant density interaction.
We discuss some implications of this result in \S~5.
Our conclusions are listed in \S~6.

\section{AFTERGLOW LIGHT CURVES IN WINDS}

\subsection{Blast Wave Hydrodynamics}

The basic model for GRB afterglow hydrodynamics involves a
relativistic blast wave expanding into the surrounding medium
(e.g., M\'esz\'aros \& Rees 1997).
During the very early evolution, the GRB ejecta play a role and
we treat the hydrodynamics of that phase in \S~3.
The later blast wave, which is dominated by the energy deposited in
the external medium, can be described by a self-similar solution
(Blandford \& McKee 1976).
The solution for expansion in a constant density medium has been widely
used in GRB studies and here we discuss some of the basic results
for a medium with density $\rho = A r^{-s}$, where $A$ is a constant.
The main problem is to determine the blast wave characteristics seen
by an external observer.

For an ultrarelativistic, adiabatic blast wave, Blandford \& McKee (1976,
their equation [69])
find that
\begin{equation}
E = {8\pi A\Gamma^2 R^{3-s}c^2\over 17-4s},
\label{a1}
\end{equation}
where $E$ is the total energy, $\Gamma$ is the Lorentz factor of the
shock front, and $R$ is the shock wave radius.
The ultrarelativistic shock condition yields the Lorentz factor
of the gas, $\gamma=\Gamma /\sqrt{2}$.
Because $E$ is constant, we have the standard result $\gamma\propto
R^{-(3-s)/2}$.
For an observer viewing the blast wave along the line of sight
to the center (not at a cosmological distance), 
consideration of emission from the blast wave at two
times in the blast wave frame yields the time in the observer's frame
(e.g., Sari 1997; Panaitescu \& M\'esz\'aros 1998; Dai \& Lu 1998):
\begin{equation}
t={R_L\over 4(4-s)\gamma_L^2 c},
\label{a2}
\end{equation}
where the subscript $L$ refers to the line of sight.
For $s=0$, we have $t=R_L/(16\gamma_L^2 c)$, as in Sari (1997).
For the wind case considered here ($s=2$), we have
$t=R_L/(8\gamma_L^2 c)$ (Dai \& Lu 1998).
The blast wave is undecelerated for $s=3$ and we obtain $t=R_L/(4\gamma_L^2 c)$;
this can also be expressed as $t=R_L/(2\Gamma_L^2 c)$, which is the standard
result for a point moving toward the observer at constant velocity.
Substitution into equation (\ref{a1}) then yields
\begin{equation}
R_L=\left[(4-s)(17-4s)Et\over 4\pi A c\right]^{1/(4-s)}.
\label{a3}
\end{equation}
We then have $R_L=(17Et/ \pi\rho_o c)^{1/4}$, where $\rho_o$ is the
ambient density, for $s=0$ and
$R_L=(9Et/ 2\pi A c)^{1/2}$ for $s=2$.

The problem with these expressions is that they apply only to
the line of sight.
In observing a burst, the emission can be dominated by emission away
from the center because the burst is observed at an earlier time
when it may have been brighter
(Waxman 1997c, Panaitescu \& M\'esz\'aros 1998; Sari 1998).
The appearance depends on the evolution of the burst, which itself
depends on the observing frequency.
Panaitescu \& M\'esz\'aros (1998) considered these issues for
wind and constant density surroundings and for radiative and
non-radiative blast waves.
The radius and Lorentz factor of the typical material that is observed
can be written as $R=\zeta R_L$ and $\gamma=\zeta^{-1/2}\gamma_L$,
where $\zeta$ is a constant and the relation between 
$R$ and $\gamma$ follows the
$\gamma\propto R^{-1/2}$ relation expected for an adiabatic blast wave in
a wind.
Panaitescu \& M\'esz\'aros (1998) find $\zeta=0.56$ at high frequencies
where $F_{\nu}\propto \nu^{-(p-1)/2}$ and
$\zeta=0.78$ at low frequencies
where $F_{\nu}\propto \nu^{1/3}$.
We thus have for a distant observer
\begin{equation}
R=1.1\times 10^{17} \left(1+z\over 2\right)^{-1/2}
E_{52}^{1/2}A_{\star}^{-1/2}t_{\rm day}^{1/2}{\rm~cm}
\label{a4}
\end{equation}
and
\begin{equation}
\gamma=5.9 \left(1+z\over 2\right)^{1/4}
E_{52}^{1/4}A_{\star}^{-1/4}t_{\rm day}^{-1/4}
\label{a5}
\end{equation}
at high frequencies,
where $E_{52}$ is the blast wave energy in units of $10^{52}$ ergs,
$t_{\rm day}$ is the observer's time in units of days,
$A=\dot M_{\rm w}/4\pi V_{\rm w}=5\times 10^{11}A_{\star}$ g cm$^{-1}$,
$\dot M_{\rm w}$ is the mass loss rate, and $V_{\rm w}$ is the wind velocity. 
The reference value of $A$ corresponds to $\dot M_{\rm w}=1\times 10^{-5}\ml$
and $V_{\rm w}=1000\kms$ (see CL for a justification of these values
in terms of the wind from a Wolf-Rayet star).
In addition, the cosmological redshift $z$ enters because of time dilation.
At low frequencies, the coefficient in equation (\ref{a4}) is increased
to $1.6 \times 10^{17}$ cm and that in equation (\ref{a5}) is reduced to 5.0.

At early times, the adiabatic assumption is expected to break down,
because all of the electron energy can be radiated by synchrotron
radiation close to the shock front.
The importance for the hydrodynamics is determined by the fraction
of the total shock energy that goes into the electrons, $\epsilon_e$.
The blast wave is radiative either if $\epsilon_e$ is close to 1 or
$\epsilon_e$ is small, but the magnetic and nucleon energies efficiently
couple to the electron energy.
In the limiting case, all of the shock power is radiated and we
have a radiative blast wave (Blandford \& McKee 1976; 
Vietri 1997; Sari et al. 1998).
For $s=2$, the blast wave radius along the line of sight is
(B\"ottcher \& Dermer 1999)
\begin{equation}
R_{rad,L}=\left(3 E_o^2 t\over 16\pi^2 A^2 c^3 \Gamma_o^2\right)^{1/3},
\end{equation}
where $E_o$ is the initial energy and $\Gamma_o$ is the initial Lorentz
factor of the shock front.
The evolution of the shock wave Lorentz factor is given by
\begin{equation}
\Gamma_{rad,L}=\left(E_o \over 12\pi A^2 c^3 \Gamma_o\right)^{1/3}t^{-1/3}.
\end{equation}
The estimated values of $\zeta$ for this case are 0.83 (low frequency)
and 0.68 (high frequency) (Panaitescu \& M\'esz\'aros 1998),
where now $R=\zeta R_{rad,L}$
and $\Gamma=\zeta^{-1}\Gamma_{rad,L}$.

The limiting radiative case is unlikely to be achieved and the actual case
is probably intermediate between the adiabatic and strongly radiative
cases.
If $\epsilon=\epsilon_e\epsilon_{rad}$ is a constant, where the synchrotron
radiative efficiency $\epsilon_{rad}$ is between 0 and 1, we have
(B\"ottcher \& Dermer 1999)
\begin{equation}
R\propto t^{(2-\epsilon)/(4-\epsilon)},\qquad 
\gamma\propto t^{-1/(4-\epsilon)}.
\label{radR}
\end{equation}
During the radiative phase, $\epsilon_{rad}\approx 1$ so that these
expressions apply to the case where $\epsilon_e$ is constant.

\subsection{Afterglow Properties with Slow Cooling}

Our treatment of afterglow light curves in winds follows the discussions
of afterglows in a constant density interstellar medium (Rees \& M\'esz\'aros
1997; Waxman 1997a,b; Sari, Piran, \& Narayan 1998; Wijers \& Galama 1999).
We presume that
 electrons are accelerated in the blast wave shock wave to a power
energy distribution, $N(\gamma)\propto \gamma^{-p}$ for $\gamma >
\gamma_m$ where $\gamma_m$ is the minimum Lorentz factor at 
the shock front.
The value of $\gamma_m$ is
\begin{equation}
\gamma_m =\left({2\over 1+X}\right)\left({m_p\over m_e}\right)
\left({p-2\over p-1}\right) \epsilon_e \gamma,
\end{equation}
where $X$ is the hydrogen mass fraction, $m_p$ and $m_e$ are the proton
and electron mass respectively, and $\epsilon_e$ is again the ratio of the
energy density in electrons to the total postshock energy density.
Our expression agrees with that of Sari et al. (1998), but differs slightly
from that of Wijers \& Galama (1999) who identify $\epsilon_e$ as the
the ratio of the
energy density in electrons to the  postshock {\it nucleon} energy density.
We take $X=0$ because we are assuming the wind is from a Wolf-Rayet star.
The electron energy for which synchrotron losses are important is 
estimated as that at which the cooling time for electrons with pitch angle
$\pi/2$ equals the expansion time:
\begin{equation}
\gamma_c={ 6\pi m_e c\over \sigma_{_T}\gamma B_{\rm{tot}}^2 
	t_{\rm{loc}} },
\label{16}
\end{equation}
where $\sigma_{_{T}}$ is the Thomson cross section, $B_{\rm{tot}}$
is the total field strength, and $t_{\rm{loc}}
=t/(1+z)$ is the time observed by an observer cosmologically local
to the burst.

The synchrotron spectrum of the afterglow can be divided into 4 power law 
sections that
are separated at 3 characteristic frequencies: the synchrotron self-absorption
frequency $\nu_A$, the characteristic frequency $\nu_m$
 emitted by electrons with
Lorentz factor $\gamma_m$, and the frequency at which synchrotron losses
become important $\nu_c$ (Sari et al. 1998).
If the frequencies are ordered $\nu_A < \nu_m < \nu_c$, the four sections
of the spectrum can be described by $F_{\nu}\propto \nu^{\beta}$ with
$\beta = 2, 1/3, -(p-1)/2, -p/2$ going from low to high frequency.
The observations of GRB 970508 on day 12 can be 
approximately represented by this
spectrum (Galama et al. 1998b) and we 
initially assume this ordering of the characteristic
frequencies in this section.
The slow cooling condition can be expressed as $\nu_m < \nu_c$ 
(Sari et al. 1998).
The peak flux, $F_{\nu,{\rm max}}$, occurs at $\nu_m$ for this case.

In any power law segment of the spectrum, the flux evolution is expected
to follow a power law $F_{\nu}\propto \nu^{\beta}t^{\alpha}$.
For a constant density medium ($s=0$), the standard result is
$\alpha=-3(p-1)/4=3\beta/2$ for $\nu_m < \nu < \nu_c$ and
$\alpha=-(3p-2)/4=(3\beta+1)/2$  for $\nu_c < \nu$.
In a wind ($s=2$), we have
$\alpha=-(3p-1)/4=(3\beta-1)/2$ for $\nu_m < \nu < \nu_c$ and
the same evolution as the $s=0$ case for $\nu_c < \nu$.
A signature of wind interaction afterglows in the non-cooling, high frequency
phase of evolution is a relatively rapid rate of decline.
The plausible assumption that most of the electron energy is
near $\gamma_m$ requires $p>2$ and $\alpha < -1.25$.
Many optical afterglows are observed to initially have $\alpha \gsim -1.3$
and so are unlikely to be in the non-cooling, wind category.
However, if the optical emission is in the cooling regime,
$\alpha=-(3p-2)/4$ as in the $s=0$ case and the requirement $p>2$
implies $\alpha < -1.0$.
More candidate afterglows are potentially in this category, but there
is the additional requirement of a moderately steep observed spectral index:
$\beta=-p/2$ so that $p>2$
implies $\beta < -1.0$.

The observed values of $F_{\nu,{\rm max}}$, $\nu_m$,  $\nu_A$, and $\nu_c$
 have been used to find the blast wave energy $E$, ambient
 density $n$, electron energy fraction $\epsilon_e$, and magnetic
 energy fraction $\epsilon_B$ in the context of $s=0$ models
 (Wijers \& Galama 1999; Granot et al. 1999).
In wind models, the constant density is replaced by the wind
density $\rho=Ar^{-2}$, where $A=\dot M_{\rm w}/4\pi V_{\rm w}=5\times 
10^{11}A_{\star}$ 
g cm$^{-1}$,
as discussed below equation (\ref{a5}).
In CL, expressions were derived for the characteristic frequencies and
$F_{\nu,{\rm max}}$ in terms of the model parameters.
We repeat the expressions for completeness:
\begin{equation}
F_{\nu,{\rm max}}= 20\left(\sqrt{1+z}-1\over \sqrt{2}-1\right)^{-2}
\left(1+z\over 2\right)^{1/2}
\left(\epsilon_B\over 0.1\right)^{1/2} E_{52}^{1/2} A_{\star}
t_{\rm day}^{-1/2} {\rm~mJy},
\label{a6}
\end{equation}
\begin{equation}
\nu_A\approx 1\times 10^{11} \left(1+z\over 2\right)^{-2/5}
\left(\epsilon_e\over 0.1\right)^{-1}
\left(\epsilon_B\over 0.1\right)^{1/5}E_{52}^{-2/5}A_{\star}^{6/5}
t_{\rm day}^{-3/5} {\rm~Hz},
\label{a7}
\end{equation}
\begin{equation}
\nu_m= 5\times 10^{12}\left(1+z\over 2\right)^{1/2}
\left(\epsilon_e\over 0.1\right)^2
\left(\epsilon_B\over 0.1\right)^{1/2}E_{52}^{1/2}
t_{\rm day}^{-3/2} {\rm~Hz},
\label{a8}
\end{equation}
\begin{equation}
\nu_c\approx 2\times 10^{12}\left(1+z\over 2\right)^{-3/2}
\left(\epsilon_B\over 0.1\right)^{-3/2}E_{52}^{1/2}A_{\star}^{-2}
t_{\rm day}^{1/2} {\rm~Hz},
\label{a9}
\end{equation}
where 
our expressions assume a flat universe with Hubble
constant $H_o=65$ km s$^{-1}$ Mpc$^{-1}$.
We also assume that the composition of the wind gas is 
hydrogen depleted and that
the electron spectral index is $p\approx 2.5$; the variation due to
different values of $p$ is less than other uncertainties.
Our expression for $\nu_c$ uses the estimate of Sari et al. (1998).
The value of $\nu_c$ is sensitive to the value of $\zeta$ discussed
in \S~2.1 ($\propto \zeta^5$); we have taken the high frequency value.
The expression for the flux here and elsewhere in the paper can be
generalized to any cosmology by replacing $(\sqrt{1+z}- 1)$ by
$d_L/(9.23{\rm~Gpc}~\sqrt{1+z})$, where
 $d_{_L}$ is the luminosity distance.
 In the case of a flat universe,
\begin{equation}
d_{_L}={2c\over H_o}(1+z-\sqrt{1+z}).
\label{21}
\end{equation}

We now invert these expressions in order to solve for the model
parameters.
The observations are taken to all refer to the same day, $t$.
Then
\begin{equation}
E \approx 3\times 10^{52} 
y^{3} x^{-1/2}
\left(t\over {\rm day}\right)^{-1/2}
\left(F_{\nu,{\rm max}}\over {\rm mJy}\right)^{3/2} 
\left(\nu_A\over 10^9 {\rm~Hz}\right)^{-5/6} 
\left(\nu_m\over 10^{12} {\rm~Hz}\right)^{-5/12} 
\left(\nu_c\over 10^{14} {\rm~Hz}\right)^{1/4} {\rm ergs},
\label{a10}
\end{equation}
\begin{equation}
A_{\star} \approx  9\times 10^{-4}
 x
\left(t\over {\rm day}\right)^{2}
\left(\nu_A\over 10^9 {\rm~Hz}\right)^{5/3} 
\left(\nu_m\over 10^{12} {\rm~Hz}\right)^{5/6} 
\left(\nu_c\over 10^{14} {\rm~Hz}\right)^{1/2},
\label{a11}
\end{equation}
\begin{equation}
\epsilon_e \approx 0.006\ 
y^{-1} x^{1/2}
\left(t\over {\rm day}\right)^{3/2}
\left(F_{\nu,{\rm max}}\over {\rm mJy}\right)^{-1/2} 
\left(\nu_A\over 10^9 {\rm~Hz}\right)^{5/6} 
\left(\nu_m\over 10^{12} {\rm~Hz}\right)^{11/12} 
\left(\nu_c\over 10^{14} {\rm~Hz}\right)^{1/4},
\label{a12}
\end{equation}
\begin{equation}
\epsilon_B \approx 1\times 10^2
y x^{-5/2}
\left(t\over {\rm day}\right)^{-5/2}
\left(F_{\nu,{\rm max}}\over {\rm mJy}\right)^{1/2} 
\left(\nu_A\over 10^9 {\rm~Hz}\right)^{-5/2} 
\left(\nu_m\over 10^{12} {\rm~Hz}\right)^{-5/4} 
\left(\nu_c\over 10^{14} {\rm~Hz}\right)^{-5/4},
\label{a13}
\end{equation}
where
$$
x={1+z\over 2}\qquad {\rm and}\qquad
y={\sqrt{2x}-1\over \sqrt{2}-1}.
$$

The afterglow light curve depends on how the break frequencies and the
peak flux, $F_{\nu,\rm max}$, evolve with time and here we consider
the light curve without cooling. As in Sari et al. (1998) for the ISM 
case, we can define critical times at which
the break frequencies pass through a fixed frequency $\nu$.
The time that $\nu_m$ crosses the observed frequency is
\begin{equation}
t_m=60\left(1+z\over 2\right)^{1/3}
\left(\epsilon_e\over 0.1\right)^{4/3}
\left(\epsilon_B\over 0.1\right)^{1/3} E_{52}^{1/3} 
 \nu_{10}^{-2/3} {\rm~days},
\label{a14}
\end{equation}
where $\nu_{10}$ is $\nu$ in units of $10^{10}$ Hz.
The time that $\nu_A$ crosses the observed frequency is
\begin{equation}
t_A=50\left(1+z\over 2\right)^{-2/3}
\left(\epsilon_e\over 0.1\right)^{-5/3}
\left(\epsilon_B\over 0.1\right)^{1/3} E_{52}^{-2/3} A_{\star}^2
 \nu_{10}^{-5/3} {\rm~days}.
\label{a15}
\end{equation}
Provided $t_A < t_m$, the light curve is as follows:
$F_{\nu}\propto R^2\propto t$ for $t < t_A$,
$F_{\nu}\propto F_{\nu,{\rm max}}\nu_m^{-1/3}\propto t^0$ for $t_A < t < t_m$,
and $F_{\nu}\propto F_{\nu,{\rm max}}\nu_m^{(p-1)/2}\propto t^{-(3p-1)/4}$
for $t_m < t$.
The condition that $t_A = t_m$ leads to a critical time and frequency:
\begin{equation}
t_{Am}=80 \left(1+z\over 2\right)\left(\epsilon_e\over 0.1\right)^{10/3}
\left(\epsilon_B\over 0.1\right)^{1/3} E_{52} A_{\star}^{-4/3} {\rm~days},
\label{a16}
\end{equation}
and 
\begin{equation}
\nu_{Am}=7\times 10^9
\left(1+z\over 2\right)^{-1}\left(\epsilon_e\over 0.1\right)^{-3}
E_{52}^{-1} A_{\star}^2 \quad {\rm~Hz}.
\label{a17}
\end{equation}
For $\nu > \nu_{Am}$, the light curve is as described above.

For $\nu < \nu_{Am}$, there is a regime where the spectrum is
characterized by $\nu_m < \nu_A$.
In general, the flux in the self-absorbed part of the spectrum
can be described by $F_{\nu}\propto R^2 \nu^2(\gamma_e/\gamma_m)$, where
$\gamma_e$ is the Lorentz factor of the electrons that are responsible
for the emission at frequency $\nu$.
Then, for $\nu<\nu_m$, $\gamma_e=\gamma_m$ and $F_{\nu}\propto t \nu^2$
as before.
For $\nu_m<\nu<\nu_A$, $(\gamma_e/\gamma_m)= (\nu/\nu_m)^{1/2}$
so that $F_{\nu}\propto t^{7/4} \nu^{5/2}$.
Above $\nu_A$, we again have
$F_{\nu}\propto t^{-(3p-1)/4} \nu^{-(p-1)/2}$; the $\nu^{1/3}$
part of the spectrum is no longer present.
The evolution of $\nu_m$ is always $\propto t^{-3/2}$, but now
$F_{\nu_m}\propto t^{-2}$.
The peak of the spectrum is at $\nu_A\propto t^{-[3(2+p)]/[2(4+p)]}$, 
and $F_{\nu_A}\propto t^{-(1+4p)/[2(4+p)]}$.
The light curve is thus described by $F_{\nu}\propto t \nu^2$
up to $t=t_m$, followed by $F_{\nu}\propto t^{7/4} \nu^{5/2}$
and $F_{\nu}\propto t^{-(3p-1)/4} \nu^{-(p-1)/2}$.
It can be seen from the estimated value of $\nu_{Am}$ that these
considerations are relevant to radio observations of afterglows.

\subsection{Fast Cooling Case}

The description of the evolution given in CL and in \S 2.2 assumes
that $\nu_m < \nu_c$.
Although synchrotron cooling is important for the high energy electrons,
it is not important for the electrons emitting near $\nu_m$ which
have most of the energy.
Before some time, $t_o$, when $\nu_m = \nu_c$, cooling of these lower
energy electrons is rapid compared to the age
 and this phase can be referred to as having
fast cooling (Sari et al. 1998).
If $\epsilon_e$ is close to 1,  
the blast wave steadily loses energy and the
hydrodynamic evolution is termed radiative.
If $\epsilon_e$ is small and the electron energy does not couple
efficiently to the
ion or magnetic energy, the hydrodynamic evolution is 
approximately adiabatic even for $t < t_o$.
We consider this second case to be the most likely and concentrate on it.
For $\nu_m > \nu_c$, the expressions for $F_{\nu,{\rm max}}$, $\nu_m$,
and $\nu_c$ (equations [\ref{a6}], [\ref{a8}], and [\ref{a9}]) remain unchanged; the maximum
flux now occurs at $\nu_c$ instead of $\nu_m$.
The spectrum can be described by $F_{\nu}\propto \nu^{\beta}$ with
$\beta =  1/3, -1/2, -p/2$ going from low to high frequency and
the breaks at $\nu_c$ and $\nu_m$, respectively (Sari et al. 1998).
However, equation (\ref{a7}) for $\nu_A$ is no longer applicable.

Equations (\ref{a8}) and (\ref{a9}) imply that
\begin{equation}
t_o=2 \left(1+z\over 2\right)
\left(\epsilon_e\over 0.1\right)
\left(\epsilon_B\over 0.1\right) A_{\star} {\rm~days}.
\label{a18}
\end{equation}
For standard parameters, the transition to slow cooling occurs at a later
time for the wind case compared to the ISM case because of the higher
densities that the shock front encounters at early times.

The afterglow light curve again depends on how the break frequencies and the
peak flux, $F_{\nu,\rm max}$, evolve with time.
The cooling time for the wind case is
\begin{equation}
t_c=2\times 10^{-5}\ \left(1+z\over 2\right)^3
\left(\epsilon_B\over 0.1\right)^3 E_{52}^{-1} 
A_{\star}^4 \nu_{10}^2 {\rm~days}.
\label{a19}
\end{equation}
The effects of cooling at a frequency $\nu$ are important for
$t<t_c$ and fast cooling can affect the blast wave evolution
up to time $t_o$, so that cooling has effects up to $t=\max(t_c,t_o)$.
The time that $\nu_m$ crosses $\nu$ is given by equation (\ref{a14}).
There are two possible orderings for the three times:
$t_m < t_o < t_c$ and $t_c < t_o < t_m$.
These cases are divided by a critical frequency,
$\nu_o=\nu_c(t_o)=\nu_m(t_o)$:
\begin{equation}
\nu_o=3\times 10^{12} \left(1+z\over 2\right)^{-1}
\left(\epsilon_e\over 0.1\right)^{1/2}
\left(\epsilon_B\over 0.1\right)^{-1} E_{52}^{1/2} 
A_{\star}^{-3/2} {\rm~Hz}.
\label{a20}
\end{equation}
When $\nu > \nu_o$, the ordering $t_m < t_o < t_c$ applies and the
evolution can be called the high-frequency light curve (cf. Sari et al. 1998
for the ISM case).
The evolution is described by $F_{\nu}\propto$ 
$t^{-1/4}\nu^{-1/2}$ ($t < t_m$),
$t^{-(3p-2)/4}\nu^{-p/2}$ ($t_m < t < t_o$),
$t^{-(3p-2)/4}\nu^{-p/2}$ ($t_o < t < t_c$), and
$t^{-(3p-1)/4}\nu^{-(p-1)/2}$ ($t_c < t$).
When $\nu < \nu_o$,  $t_c < t_o < t_m$ applies and we have
 the low-frequency light curve:
$F_{\nu}\propto$ 
$t^{-1/4}\nu^{-1/2}$ ($t < t_c$),
$t^{-2/3}\nu^{1/3}$ ($t_c < t < t_o$),
$t^{0}\nu^{1/3}$ ($t_o < t < t_m$), and
$t^{-(3p-1)/4}\nu^{-(p-1)/2}$ ($t_m < t$).
These light curves are distinct from the ISM case, especially
because of the early importance of cooling.

Self-absorption is important throughout the cooling regime ($t<t_o$)
if $t_A >t_o$.  From equations (\ref{a16}) 
(which assumes the low-frequency light curve)
and (\ref{a18}), the frequency at which $t_A =t_o$ is given by
\begin{equation}
\nu_{Ao}=8\times 10^{10}\left(1+z\over 2\right)^{-1}
\left(\epsilon_e\over 0.1\right)^{-8/5}
\left(\epsilon_B\over 0.1\right)^{-2/5} E_{52}^{-2/5} 
A_{\star}^{3/5} {\rm~Hz}.
\label{b20}
\end{equation}
With typical parameters, $\nu_{Ao} < \nu_o$ so that the use of 
equation (\ref{a16})
is justified.
At radio wavelengths, the synchrotron emission is self-absorbed during
the fast cooling period.
At optical and X-ray wavelengths, self-absorption is important only during
the early phases of the fast cooling period.

The effects of self-absorption during the fast cooling period
can be found in the same way as
described above.
The frequencies $\nu_m$ and $\nu_c$ are again given by the
standard expressions (equations [\ref{a8}] and [\ref{a9}]), and the behavior of
$\nu_A$ can be found by considering the complete spectral evolution.
For the cases $\nu_A<\nu_c<\nu_m$ and $\nu_c<\nu_A<\nu_m$, the
high frequency properties are the same.
For $\nu>\nu_m$, $F_{\nu}\propto t^{-(3p-2)/4}\nu^{-p/2}$ and
for $\nu$ below $\nu_m$, we have
$F_{\nu}\propto t^{-1/4}\nu^{-1/2}$.
We also have $F_{\nu_m}\propto t^{1/2}$.
In the lowest frequency range, where self-absorption is important,
the electrons responsible for the emission have a Lorentz factor
corresponding to peak emission at $\nu_c$.
The evolution is thus given by $F_{\nu}\propto R^2 (\nu_c/\nu_m)^{1/2}
\propto t^2 \nu^2$.
For $\nu_A<\nu_c$, it is now straightforward to show that
$\nu_A\propto t^{-8/5}$, $F_{\nu_A}\propto t^{-6/5}$,
$F_{\nu_c}\propto t^{-1/2}$, and $F_{\nu}\propto t^{-2/3}\nu^{1/3}$
for $\nu_A<\nu<\nu_c$.
For $\nu_c<\nu_A$, we have
$\nu_A\propto t^{-2/3}$, $F_{\nu_c}\propto t^{3}$,
$F_{\nu_a}\propto t^{1/12}$, and $F_{\nu}\propto t^{7/4}\nu^{5/2}$
for $\nu_c<\nu<\nu_A$.

\subsection{Light Curves}

The consideration of complete light curves for relativistic, spherical
expansion involves the combination of both cooling and adiabatic
evolution.
At a particular frequency, cooling effects are important up to
a time $t=\max(t_o,t_c)$.
The relevant transition times for a light curve are $t_A$, $t_o$, 
$t_c$, and $t_m$.
These four transition times allow many possible light curves, but
our previous discussion allows us to limit the possibilities, on the
assumption that the blast wave conditions are not far from the
typical conditions that we have chosen.
In particular, equations (\ref{a16}) and (\ref{a17}) 
show that we expect $t_A=t_m$ at a
relatively late time and low frequency.
We thus expect $t_A<t_m$ at higher frequencies, including all those
in the high frequency case with regard to radiative cooling.

The orderings of the transition times of interest are thus:
A) $t_A<t_m<t_o<t_c$;
B) $t_A<t_c<t_o<t_m$;
C) $t_c<t_A<t_o<t_m$;
D) $t_c<t_o<t_A<t_m$; and
E) $t_c<t_o<t_m<t_A$, where the listing is
from high frequency to low frequency light curves.
The frequency dividing light curve A from B is $\nu_o$ (equation [\ref{a20}]),
that dividing C from D is $\nu_{Ao}$ (equation [\ref{b20}]), and
that dividing D from E is $\nu_{Am}$ (equation [\ref{a17}]).
The frequency dividing curve B from C is that at which $t_c=t_A$.
By referring to the previous sections, the light curves for each
case can be constructed.
They are:
A) $F_{\nu}\propto t^{7/4}\nu^{5/2}$ $(t<t_A)$, 
$t^{-1/4}\nu^{-1/2}$ $(t_A,t_m)$,
$t^{-(3p-2)/4}\nu^{-p/2}$ $(t_m,t_o)$, $t^{-(3p-2)/4}\nu^{-p/2}$ $(t_o,t_c)$,
$t^{-(3p-1)/4}\nu^{-(p-1)/2}$ $(t_c <t)$;
B) $F_{\nu}\propto t^{7/4}\nu^{5/2}$ $(t<t_A)$, 
$t^{-1/4}\nu^{-1/2}$ $(t_A,t_c)$,
$t^{-2/3}\nu^{1/3}$ $(t_c,t_o)$, $t^{0}\nu^{1/3}$ $(t_o,t_m)$,
$t^{-(3p-1)/4}\nu^{-(p-1)/2}$ $(t_m <t)$;
C) $F_{\nu}\propto t^{7/4}\nu^{5/2}$ $(t<t_c)$, 
$t^{2}\nu^{2}$ $(t_c,t_A)$,
$t^{-2/3}\nu^{1/3}$ $(t_A,t_o)$, $t^{0}\nu^{1/3}$ $(t_o,t_m)$,
$t^{-(3p-1)/4}\nu^{-(p-1)/2}$ $(t_m <t)$;
D) $F_{\nu}\propto t^{7/4}\nu^{5/2}$ $(t<t_c)$, 
$t^{2}\nu^{2}$ $(t_c,t_o)$,
$t\nu^{2}$ $(t_o,t_A)$, $t^{0}\nu^{1/3}$ $(t_A,t_m)$,
$t^{-(3p-1)/4}\nu^{-(p-1)/2}$ $(t_m <t)$;
E) $F_{\nu}\propto t^{7/4}\nu^{5/2}$ $(t<t_c)$, 
$t^{2}\nu^{2}$ $(t_c,t_o)$,
$t\nu^{2}$ $(t_o,t_m)$, $t^{7/4}\nu^{5/2}$ $(t_m,t_A)$,
$t^{-(3p-1)/4}\nu^{-(p-1)/2}$ $(t_A <t)$.
The times in parentheses give the time range for that particular
section of the light curve.
The light curves are illustrated in Fig. 1.
In the figure, $t_A$ is omitted from curve A and $t_c$ from
curves D and E because they occur at very early times.

For our typical parameters, both X-ray and optical light curves
are of type A.
It can be seen from equation (\ref{a14}) that $t_m$ is typically in the
10's of seconds range at these high frequencies.
The self-absorbed part of the light curve occurs considerably
earlier, when the shock evolution is still affected by the
energy deposition
 and our model is not applicable.
The radio light curve is typically of type D.
The light curve properties are like those described
by CL, except we have now found that a steeper rise is
expected for $t<t_o$.

For the above light curves, we have assumed that the hydrodynamic
evolution is described by an adiabatic blast wave.
For $t<t_o$, the energy loss by electrons can affect the evolution.
In \S~2.1, we gave the expected evolution for a fully radiative
blast wave.
In this case for a $\rho\propto r^{-2}$ ambient medium,
standard expressions  (Sari et al. 1998;
Wijers \& Galama 1999) yield $\nu_c\propto t^{1/3}$,
$\nu_m\propto t^{-5/3}$, and $F_{\nu,{\rm max}}\propto t^{-2/3}$.
We thus have 
$F_{\nu}=F_{\nu,{\rm max}}(\nu/\nu_c)^{1/3}\propto
t^{-7/9}\nu^{1/3}$ for $\nu_A<\nu<\nu_c$,
$F_{\nu}=F_{\nu,{\rm max}}(\nu/\nu_c)^{-1/2}\propto
t^{-1/2}\nu^{-1/2}$ for $\nu_c<\nu<\nu_m$, and
$F_{\nu}=F_{\nu_m}(\nu/\nu_m)^{-p/2}\propto
t^{-(5p-2)/6}\nu^{-p/2}$ for $\nu_m<\nu$.
The evolution at optical and X-ray wavelengths is most likely
to be in the $\nu_m<\nu$ regime; for $p=(2.0,2.5,3.0)$, the
evolution of a radiative blast wave is given by
$F_{\nu}\propto t^{(-1.33,-1.75,-2.17)}$, as compared to
$F_{\nu}\propto t^{(-1.00,-1.38,-1.75)}$ for an adiabatic
blast wave.
There is a significant steepening of the light curve for
the radiative case.
The radio light curve is likely to be in the self-absorbed regime.
For $\nu<\nu_A,\nu_c$, we have $F_{\nu}\propto R^2 \nu^2(\nu_c/\nu_m)^{1/2}
\propto t^{5/3}\nu^2$.

The actual situation is likely to be intermediate between
adiabatic and radiative evolution of the blast wave and we
use equations (\ref{radR}) to find the time dependences in that case.
We find $\nu_c\propto t^{(2-\epsilon)/(4-\epsilon)}$,
$\nu_m\propto t^{-(6-\epsilon)/(4-\epsilon)}$,
and $F_{\nu,max}\propto t^{-2/(4-\epsilon)}$.
The resulting flux evolution for $\nu>\nu_m$ is 
$F_{\nu}\propto t^{-[(6-\epsilon)p-2(2-\epsilon)]/[2(4-\epsilon)]}\nu^{-p/2}$.
Thus, for $\epsilon=0.1$, we have
$F_{\nu}\propto t^{(-1.03,-1.40,-1.78)}$ for
$p=(2.0,2.5,3.0)$, and for $\epsilon=1/3$, we have
$F_{\nu}\propto t^{(-1.09,-1.48,-1.86)}$.
It can be can that the change from the adiabatic case is small
for $\epsilon=0.1$, but that it becomes significant for $\epsilon=1/3$.
The radio evolution is likely to be characterized by $\nu<\nu_A$,
which is below $\nu_c$ and $\nu_m$.
We then have $F_{\nu}\propto t^{(8-3\epsilon)/(4-\epsilon)}\nu^2$.

\subsection{Transition to Nonrelativistic Evolution}

The transition to nonrelativistic evolution has been discussed 
 for the ISM, constant density case by Wijers et al. (1997) and
Waxman, Kulkarni, \& Frail (1998).
For an ambient density  $\rho= Ar^{-2}$, 
a straightforward estimate of the transition radius 
and time is when the shock has swept up a mass equivalent to the
rest mass of the explosion:
\begin{equation}
r_{\rm NR}={E\over 4\pi Ac^2}
=1.8\times 10^{18}{E_{52}\over A_{\star}} {\rm~cm}
\label{a21}
\end{equation}
and
\begin{equation}
t_{\rm NR}={r_{\rm NR}\over c}=1.9 {(1+z)}{E_{52}\over A_{\star}} {\rm~yr}.
\label{a22}
\end{equation}
   From equation (\ref{a5}), the relativistic blast wave shock 
   Lorentz factor at this time
is $\Gamma= 1.38$ (high frequency observation) or
$\Gamma= 1.17$ (low frequency observation).

The nonrelativistic
blast wave in a $\rho=Ar^{-2}$ medium 
is particularly simple for adiabatic index
$\gamma_a=5/3$ (e.g., Sedov 1959).
The shock radius is given by
\begin{equation}
r_s=\left(3E\over 2\pi A\right)^{1/3}t^{2/3}
\label{a23}
\end{equation}
and the postshock profiles are $\rho=\rho_s(r/r_s)$,
$v=v_{gs}(r/r_s)$, and $p=p_s(r/r_s)^{3}$, where the subscript $s$
refers to the value at the shock front and $v_{gs}$ is the gas
velocity at the shock front.
The blast wave expansion can be described by
$r_s=\xi_o(E/A)^{1/3}t^{2/3}$, where $\xi_o=0.78$ if $\gamma_a=5/3$ 
(nonrelativistic fluid) and
$\xi_o=0.64$ if $\gamma_a=4/3$ (relativistic fluid).
The $\gamma_a=5/3$ evolution yields $\dot r_s/c =1.2$ at the time
$t_{\rm NR}$.
The relativistic and nonrelativistic expansion approximately cross
at $t_{\rm NR}$, showing that this is a reasonable estimate of the
transition time.
Equation (\ref{a22}) shows that the transition to nonrelativistic evolution in
a wind is late for typical parameters;
the expected value of $t_{\rm NR}$ is longer than the typical duration
of observations of GRB afterglows.

In considering the evolution during the nonrelativistic regime, the
treatment is similar to that in the relativistic regime provided
$\gamma_m \gsim 1$.
We have $R\propto t^{2/3}$, shock velocity $v_{sh}\propto t^{-1/3}$,
$\rho_1 v_{sh}^2\propto t^{-2}$, $B\propto t^{-1}$, and
$\gamma_m\propto v_{sh}^2\propto t^{-2/3}$, so that
$\nu_m\propto \gamma_m^2 B\propto t^{-7/3}$ and $F_{\nu_m}
\propto NB\propto Rt^{-1}\propto t^{-1/3}$, where 
$\rho_1$ is the preshock density and $N$ is the total
number of radiating particles.
Then, $F_{\nu}\propto F_{\nu_m} \nu_m^{-1/3}\propto t^{4/9}$ ($\nu < \nu_m$)
and $F_{\nu}\propto F_{\nu_m} \nu_m^{(p-1)/2}\propto t^{(5-7p)/6}$ 
($\nu > \nu_m$).
For example, $p=3$ yields $F_{\nu}\propto t^{-8/3}$ in the nonrelativistic
case for $\nu > \nu_m$, but $F_{\nu}\propto t^{-2}$ in the relativistic
case.
As in interstellar interaction, the light curve steepens after the
transition to nonrelativistic flow.

\subsection{Jets}

In most models for GRBs, the burst energy is initially deposited
in a relativistic jet.
As long as the jet is highly relativistic, the observed features
should be reproduced by spherical models, but as the shocked jet
slows there are effects on the observed light curve (as first
discussed by Rhoads 1997).
If $\theta_o$ is the angular width of the jet, the edge of the jet
becomes visible when $\gamma\approx \theta_o^{-1}$ (Panaitescu 
\& M\'esz\'aros 1999).
The other important effect is that the slowed jet is able to
expand laterally.
Rhoads (1997, 1999) estimates that this will occur when
$\gamma\approx \theta_o^{-1}/\sqrt{3}$, but Sari, Piran, \& Halpern (1999)
argue that the transition occurs when $\gamma\approx \theta_o^{-1}$.
The issue is the speed of sideways expansion and 2-dimensional
numerical simulations may be needed to provide reliable results.
The important point is that during the spreading phase, there is exponential
slowing down of the forward shock front (Rhoads 1997).
Rhoads (1999) has discussed this phenomenon in detail for interstellar
interaction, but suggests that it also occurs for expansion in a wind.

Because jet effects are expected when $\gamma\approx \theta_o^{-1}$,
equation (\ref{a5}) can be used to find the time
\begin{equation}
t_{\rm jet}=2\left(1+z\over 2\right)\left(\theta_o\over 0.2\right)^4
E_{52} A_{\star}^{-1}{\rm ~day}
\label{a24}
\end{equation}
when  jet effects become important.
This expression is appropriate for high frequencies; at low 
frequencies (radio),
the coefficient becomes 1 day.
A steepening of the afterglow light curve is expected at $t_{\rm jet}$.
Once lateral spreading of the jet becomes important, the rapid
slowing implies that   $R$ becomes
essentially constant with time and this defines the hydrodynamic evolution
(Sari et al. 1999).
The general nature of the slow expansion is shown by the conservation
of energy, $E\approx c^2\rho_1 R^3\gamma^2\Omega$, where $\rho_1$ is
the preshock density and $\Omega$ is the solid angle of the jet.
During the jet spreading phase $\Omega\propto \gamma^{-2}$
(Rhoads 1997, 1999; Sari et al. 1999) and $E\approx c^2\rho_1 R^3$,
so that $R$ must be a weaker function of $\gamma$ than any power
law, independent of the density profile.\footnote{Re'em Sari pointed out
this argument.}
Thus, both
ISM and wind interactors should show the same emission properties
during the spreading phase.
We expect 
$F_{\nu}\propto$ 
$t^{0}\nu^{2}$ ($\nu < \nu_A$),
$t^{-1/3}\nu^{1/3}$ ($\nu_A < \nu < \nu_m$),
$t^{-p}\nu^{-(p-1)/2}$ ($\nu_m < \nu < \nu_c$), and
$t^{-p}\nu^{-p/2}$ ($\nu_c < \nu$) (Sari et al. 1999).

\section{PROMPT, REVERSE SHOCK EMISSION}

An initially relativistically hot fireball associated with a GRB
cools as it expands, 
forming a cold shell of material coasting at an ultrarelativistic 
speed when most of the internal energy is converted into the bulk 
kinetic energy. When the coasting shell runs into an ambient 
medium, the bulk kinetic energy is gradually released back into 
the internal energy, which gives rise to  prompt emission. 
Prompt emission has been discussed by Sari \& Piran (1999a,b)
and M\'esz\'aros \& Rees (1999) in the context of GRB 990123, both
assuming a constant ambient density suitable for the interstellar
medium. In \S~3.1, we extend their discussion of prompt 
emission to the case of a stellar wind as the ambient medium. 
A comparison with the ISM case is presented in \S~3.2.

\subsection{Wind Interaction Case}
 
The structure of the coasting shell is largely unknown, and is 
determined by the way that mass and energy are injected from 
the central engine. It is likely to be inhomogeneous, since bursts 
of gamma rays are thought to come from one part of the shell
running into another. We shall,
however, ignore such inhomogeneities and adopt uniform 
distributions for both the shell density and Lorentz factor 
(denoted by $\gamma_{\rm{sh}}$) for simplicity. 

As is well known, the shell interacts with the ambient medium via 
two shocks: a forward shock and a reverse shock. The forward shock 
runs forward into the ambient medium, whereas the reverse shock 
sweeps up the shell material. The shocked ambient and shell
materials are in pressure balance and are separated by a contact 
continuity. Since the thickness of the shocked region is expected
to be much smaller than its radius, the whole interaction region can 
be treated as planar, as done by Katz (1994) and Sari \& Piran
(1995). We will follow these authors in determining the shock
properties in a planar geometry, assuming in addition that only
a small fraction of the internal energy of the shocked materials 
is radiated away by electrons. 

There are four regions of distinct properties: the unshocked
ambient medium (denoted by ``1''), the shocked ambient 
medium (denoted by ``2''), the shocked shell material (denoted
by ``3'') and the unshocked coasting shell (denoted by ``4''). 
We assume that 
the shocked
materials in regions 2 and 3 are uniform and move together, and thus
share a common bulk Lorentz factor.  From relativistic 
shock jump conditions, we find that the Lorentz factor is
\begin{equation}
\gamma_{12}={\xi^{1/4} \gamma_{\rm{sh}}^{1/2}\over \sqrt{2}},
\label{1}
\end{equation}
(measured relative to the nearly static ambient medium, region 1) 
where  
$\xi\equiv \rho_4/\rho_1$ is the ratio of proper mass densities 
in the unshocked shell (region 4) and the ambient medium (region 
1). The proper mass density in the ambient medium is given by
\begin{equation}
\rho_1={ {\dot M}_{\rm{w}}\over 4\pi R^2 V_{\rm{w}} },
\label{2}
\end{equation}
where ${\dot M}_{\rm{w}}$ and $V_{\rm{w}}$ are the mass loss rate
and the velocity of the ambient wind, and $R$ the spherical 
radius. The proper density of the unshocked shell is given by
\begin{equation}
\rho_4={M_{\rm{sh}}\over 4\pi R^2 \gamma_{\rm{sh}}\Delta},
\label{3}
\end{equation}
where $M_{\rm{sh}}$ and $\Delta$ are the proper mass and the width 
(measured in the frame at rest with respect to the origin and
the ambient medium and cosmologically local to the burst) 
of the initial shell. The proper mass is related 
to the total initial kinetic energy $E_0$ of the coasting shell 
through
\begin{equation}
M_{\rm{sh}}={E_0\over \gamma_{\rm{sh}} c^2},
\label{4}
\end{equation}
where $c$ is the speed of light. From equations (\ref{2})--(\ref{4}), we have
\begin{equation}
\xi={E_0 V_{\rm{w}}\over {\dot M}_{\rm{w}}\gamma_{\rm{sh}}^2 
c^2\Delta},
\label{5}
\end{equation}
which is independent of radius. Therefore, the shocked materials 
move at a constant speed according to equation (\ref{1}). This fact 
simplifies our discussion of their emission properties.

The initial shell width $\Delta$ is unknown a priori. It is 
related to the time $T_{\rm{cr}}$ in the frame at rest with
respect to the origin and cosmologically local to the burst 
for the reverse shock to cross the entire shell through
\begin{equation}
T_{\rm{cr}}={\gamma_{12}^2\Delta\over c}.
\label{new1}
\end{equation}
The radiation emitted at this time by the shocked shell material
along the line of sight is received by an observer on Earth at 
different times, ranging
from $(1+z)\Delta/(2c)$ (for radiation emitted near the contact 
discontinuity) to $(1+z)\Delta/c$ (for radiation emitted 
immediately behind the reverse shock). The timescale, $\Delta/c$,
therefore characterizes the duration of prompt emission (see
Sari \& Piran 1999b).
We assume that the duration is comparable to that of the GRB itself,
which is typically 10's of seconds for the sources with observed 
afterglows. That is, the shell needs to have a 
width of tens of light seconds or more. We shall 
therefore scale $\Delta$ by 10 light seconds, and denote the scaled 
width by $\Delta_{10}$. Scaling other quantities with their 
typical values, we have
\begin{equation}
\xi=5.9 {E_{52} \over A_{\star}\Delta_{10}\gamma_3^2},
\label{7}
\end{equation}
where 
$\gamma_3$ is the unshocked shell Lorentz factor 
$\gamma_{\rm{sh}}$ divided by $10^3$ (not to be 
mistaken as the Lorentz factor of region 3).
In deriving equation (\ref{new1}), we have 
assumed that $\xi\ll \gamma_{\rm{sh}}^2$, which is true for
typical parameters. Substituting equation (\ref{7}) into equation (\ref{1}),
we finally have
\begin{equation}
\gamma_{12}=35 {E_{52}^{1/4}\over A_{\star}^{1/4}
\Delta_{10}^{1/4} },
\label{8}
\end{equation}
which is much greater than unity unless $\Delta_{10}$, the most
uncertain parameter in the above expression, is unreasonably
large.

The shocked ambient medium and the shocked shell material have not
only the same bulk Lorentz factor, but also the same internal
energy density because of pressure balance across the contact discontinuity. 
Relativistic shock jump conditions yield  
\begin{equation}
e_3=e_2=2\gamma_{\rm{sh}}\rho_1 c^2\xi^{1/2},
\label{9}
\end{equation}
where the mass density of the ambient wind is given by equation (\ref{2}),
which can be rewritten into 
\begin{equation}
\rho_1={{\dot M}_{\rm{w}}\over 4\pi V_{\rm{w}} R^2}
={{\dot M}_{\rm{w}}\over 4\pi V_{\rm{w}} c^2 T^2},
\label{10}
\end{equation}
with the time $T$ measured in the frame at rest with respect to
the origin and cosmologically local to the burst. 
As usual, we assume a (small) fraction $\epsilon_e$ 
of the internal energy of the shocked matter goes into radiating
electrons with a power-law energy distribution. Adopting a 
constant power-law index of $p$, the minimum electron Lorentz
factors in the two shocked regions are
\begin{equation}
\gamma_{m2}=\left({p-2\over p-1}\right)\left({2\over 1+X}\right) 
	\left({m_p\over m_e}\right)
	{\epsilon_{e2}\gamma_{\rm{sh}}^{1/2}\xi^{1/4}\over \sqrt{2}},
\label{11}
\end{equation}
and 
\begin{equation}
\gamma_{m3}=\left({p-2\over p-1}\right)\left({2\over 1+X}\right)
	\left({m_p\over m_e}\right)
 	{\epsilon_{e3}\gamma_{\rm{sh}}^{1/2}\over \sqrt{2}\xi^{1/4}},
\label{12}
\end{equation}
where $m_p$ and $m_e$ are the mass of protons and electrons, respectively,
and $X$ is the fractional abundance of hydrogen, which is
close to zero for  hydrogen depleted Wolf-Rayet winds. In addition, we 
assume that a constant fraction $\epsilon_B$ of the internal 
energy goes into the magnetic field in both the shocked ambient
medium and the shocked shell material, which yields a total
field strength of
\begin{equation}
B_{\rm{tot}}=(8\pi\epsilon_B e_2)^{1/2}=\left({4\epsilon_B\gamma
_{\rm{sh}}{\dot M}_{\rm{w}}\xi^{1/2}\over V_{\rm{w}}}\right)^{1/2} 
{1\over T},
\label{13}
\end{equation}
in regions 2 and 3. 

To determine the shock emission properties, we adopt the formalism 
of Sari, Piran \& Narayan (1998), including cosmological
corrections. Let us first consider the reverse 
shock (region 3). Ignoring the synchrotron self-absorption for the 
moment, there are two characteristic frequencies that we need to
determine: the ``typical'' frequency $\nu_{m3}$ corresponding to
the minimum Lorentz factor $\gamma_{m3}$ and the cooling frequency
$\nu_c$. In the observer's frame, we find 
\begin{equation}
\nu_{m3}=\left({1\over 1+z}\right) {\gamma_{12} 
	\gamma_{m3}^2\ e \ B_{\rm{tot}}\over 2\pi m_e c}
	=9.8\times 10^{18} \left({3p-6\over p-1}\right)^2
	\left({1\over 1+X}\right)^2\left({\epsilon_{e3}\over 0.1}\right)^2
	\left({\epsilon_B\over 0.1}\right)^{1/2}{A_{\star}\Delta_{10}^{1/2}
	\gamma_{3}^2\over E_{52}^{1/2} \ t } \ \ 
	{\rm Hz},
\label{14}
\end{equation}
where $e$ is the charge of electron and the factor $(3p-6)/(p-1)$ 
equals unity for 
$p=2.5$. We have related the observer's time $t$ to the time $T$ in the 
frame at rest with respect to the origin through \footnote
{Strictly speaking, the radiation emitted at time $T$ from the shocked 
material along the line of sight between the forward and reverse shocks 
arrives at the observer at different times. The observed times are $(1+z)T/
\gamma_{12}^2$, $(1+z)T/(2\gamma_{12}^2)$, and $(1+z)T/(4\gamma_{12}^2)$
for the radiation emitted at time $T$ from the reverse shock front, the 
contact continuity, and the forward shock front, respectively. We pick
$t=(1+z)T/(2\gamma_{12}^2)$ as a compromise between the reverse shock and 
the forward shock, since the emission from both regions will be 
computed.} 
\begin{equation}
t ={1+z\over 2}{T\over\gamma_{12}^2}.
\label{15}
\end{equation}
The ``typical'' frequency $\nu_{m3}$ shown in equation (\ref{14}) is to 
be compared with the cooling frequency $\nu_c$ corresponding to 
the cooling Lorentz factor given by equation (\ref{16}) with $\gamma=\gamma_{12}$.
The cooling frequency observed on Earth is then 
\begin{equation}
\nu_c=\left({1\over 1+z}\right){\gamma_{12}\gamma_c^2
	e B_{\rm{tot}}\over 2\pi m_e c}
	=3.6\times 10^8\left({2\over 1+z}\right)^2
	\left({\epsilon_B\over 0.1}\right)^{-3/2} 
	{ E_{52}^{1/2} \ t \over A_{\star}^2 \Delta_{10}^{1/2} } \ \ \rm{Hz}.
\label{17}
\end{equation}
Clearly, the cooling frequency is much lower than the ``typical'' 
frequency for reasonable parameters,
 indicating that most 
radiating electrons cool quickly down to the cooling Lorentz
factor $\gamma_c$. In other words, the shocked shell material
is in the fast cooling regime of Sari, Piran \& Narayan (1998). 
The emitted flux density therefore peaks at $\nu_c$ instead of $\nu_{m3}$, 
with a peak value given approximately by 
\begin{equation}
F_{\nu_c,3}={N_{e,3}(1+z)P_{\nu,\rm{max}}\over	4\pi d_{_L}^2},
\label{18}
\end{equation}
where the number of radiating electrons $N_{e,3}$ in the shocked shell 
region increases linearly with time as 
\begin{equation}
N_{e,3}=\left({2\over 1+z}\right)\left({1+X\over 2}\right){E_0 t
	\over \gamma_{\rm{sh}}m_p c\Delta}.
\label{19}
\end{equation}
The peak spectral power $P_{\nu,\rm{max}}$ is given roughly by
\begin{equation}
P_{\nu,\rm{max}}={m_e c^2\sigma_{_T} \gamma_{12}B_{\rm{tot}}\over 3 e}.
\label{20}
\end{equation}
 Substituting equations (\ref{21}), (\ref{19}),
and (\ref{20}) into
equation (\ref{18}), we arrive at an approximate
peak flux at the cooling frequency of
\begin{equation}
F_{\nu_c,3}=16 (1+X) \left({1+z\over 2}\right) \left({ 2-\sqrt{2} \over 
	1+z-\sqrt{1+z} }\right)^2 \left({\epsilon_B\over 0.1}\right)
	^{1/2} {E_{52} A_{\star}^{1/2}\over \gamma_{3}\Delta_{10}
	 } \ \ \rm{Jy},
\label{22}
\end{equation}
which is independent of time.

With the two characteristic frequencies ($\nu_{m3}$ and $\nu_c$) and
the peak flux ($F_{\nu_c,3}$) thus determined, we can now obtain 
the flux at any
given frequency. We are particularly interested in the optical 
prompt emission, say in the R-band at $\nu_{_R}=4.5\times 10^{14}$ Hz,
a frequency well above the cooling frequency $\nu_c$ but below
the typical frequency $\nu_{m3}$ for typical 
parameters. In this spectral regime, we have (cf. Sari et al. 1998)
$$
F_{\nu_{_R},3}=\left({\nu_{_R}\over \nu_c}\right)^{-1/2}F_{\nu_c,3}
$$
\begin{equation}
	=14 (1+X) \left({2-\sqrt{2}\over 1+z-\sqrt{1+z} }\right)^2
	\left({\epsilon_B\over 0.1}\right)^{-1/4}
	{E_{52}^{5/4}t^{1/2} \over A_\star^{1/2} \gamma_{3}
	\Delta_{10}^{5/4} } \ \ \rm{mJy}.
\label{23}
\end{equation}
Note that in this case the optical flux increases with the square root 
of the observer's time, and a maximum is reached when the reverse shock
crosses the inner edge of the freely coasting shell at the time
\begin{equation}
t_{\rm cr}=10 {1+z\over 2}\Delta_{10} \ \ {\rm s}.
\label{24}
\end{equation}
The corresponding maximum flux is
\begin{equation}
F_{\nu_{_R},3}^{\rm{max}}=46 (1+X) \left({1+z\over 2}\right)^{1/2}
	\left({2-\sqrt{2}\over 1+z-\sqrt{1+z} }\right)^2
	\left({\epsilon_B\over 0.1}\right)^{-1/4}
	{E_{52}^{5/4}
	\over A_\star^{1/2} \gamma_{3}\Delta_{10}^{3/4} } \ \ 
	\rm{mJy}.
\label{25}
\end{equation}
Interestingly, the peak flux is not sensitive to the magnetic energy 
fraction $\epsilon_B$, one of the  most uncertain parameters.
Note that a flux density of 46 mJy in the R-band corresponds to a magnitude 
of 12. Therefore, it is difficult to create a 9$^{\rm{th}}$-magnitude 
optical flash that lasts for several tens of seconds, as observed in 
GRB 990123 (Akerlof et al. 1999), in the reverse shock if the 
explosion occurs in a typical Wolf-Rayet wind,  unless the 
parameters are far from typical, e.g., a much higher explosion energy 
than $10^{52}$ ergs (see \S~4.3 for a discussion).

For completeness, we now discuss briefly the prompt, optical emission 
from the forward shock. The ambient medium shocked by the
forward shock has the same cooling frequency $\nu_c$ as the shell 
material shocked by the reverse shock, but a substantially
higher typical frequency of
$$
\nu_{m2}=\left({1\over 1+z}\right){\gamma_{12} 
	\gamma_{m2}^2 e B_{\rm{tot}}\over 2\pi m_e c}
$$
\begin{equation}
	=5.7\times 10^{19} \left({3p-6\over p-1}\right)^2
	\left({1\over 1+X}\right)^2\left({\epsilon_{e2}\over 0.1}\right)^2
	\left({\epsilon_B\over 0.1}\right)^{1/2}
	{E_{52}^{1/2}
	\over \Delta_{10}^{1/2}\ t } \ \ 
	\rm{Hz},
\label{27}
\end{equation}
where equation (\ref{11}) is used to eliminate $\gamma_{m2}$. For typical
parameters, we have $\nu_c < \nu_{_R}<\nu_{m2}$, and the flux in 
the R-band is given by 
\begin{equation}
F_{\nu_{_R},2}=\left({\nu_{_R}\over\nu_c}\right)^{-1/2}
	{N_{e,2}(1+z)P_{\nu,\rm{max}}\over 4\pi d_{_L}^2},
\label{28}
\end{equation}
where the number of radiating electrons $N_{e,2}$ in the shocked ambient
medium swept up by the forward shock increases linearly with time as
\begin{equation}
N_{e,2}=\left({1+X\over 2}\right)\left({2\over 1+z}\right)
	{ {\dot M}_{\rm{w}} c\gamma_{12}^2 \ t
	\over m_p V_{\rm{w}} }.
\label{29}
\end{equation}
Compared with the R-band flux for the reverse shock given in equation 
(\ref{23}), the flux in the forward shock is down by a factor of
\begin{equation}
{F_{\nu_{_R},2}\over F_{\nu_{_R},3}}={N_{e,2}\over N_{e,3}}
	={1\over 2\xi^{1/2}}={1\over 4.8} {A_{\star}^{1/2}
	\Delta_{10}^{1/2}\gamma_{3}\over E_{52}^{1/2}
	 }.
\label{30}
\end{equation}
Therefore, the optical flash is typically dominated by the reverse 
shock, as in the previously studied case of constant-density 
ambient medium. 

After the coasting shell is completely shocked, the reverse shock
front disappears, and there is no more shell kinetic energy left 
to drive the forward shock. Instead of maintaining a constant 
Lorentz factor, the shocked region slows down with time as more 
ambient medium is swept up. 
The forward shock front begins to evolve as described in \S~2.
For the material in the reverse shock region, we still have $\nu_c
<\nu_{m3}$ at the time that the reverse shock disappears.
The electrons in this region thus rapidly cool and do not contribute
to the emission, which is dominated by the forward shock region.

Synchrotron self-absorption may reduce our estimate of the optical
prompt emission. A simple way to gauge this effect is to 
estimate the maximal flux emitted by the shocked shell material 
as a black body 
\begin{equation}
F_{\nu,\rm{bb}}\approx (1+z)^3 \pi\left({R_\perp\over d_{_L}}\right)^2
	\left({2\nu^2\over c^2}\right) k T_{\rm{eff}},
\label{40}
\end{equation}
(Sari \& Piran 1999b) with the observed size $R_\perp$ given roughly by
\begin{equation}
R_\perp\approx 2 \gamma_{12} c t,
\label{41}
\end{equation}
and the effective temperature by
\begin{equation}
kT_{\rm eff}\approx \gamma_{12}\gamma_{\nu} m_e c^2/3,
\label{42}
\end{equation}
where $\gamma_{\nu}$ is the Lorentz factor of the electrons that radiate at 
the frequency $\nu$, and is given by $[2\pi(1+z)m_ec\nu/(\gamma_{12}e
B_{\rm{tot}})]^{1/2}$. 
Substituting equations (\ref{41}) and (\ref{42}) into equation (\ref{40}), 
we have
\begin{equation}
F_{\nu,\rm{bb}}\approx 2.0 \left({1+z\over 2}\right)^3
	\left({2-\sqrt{2}\over 1+z-\sqrt{1+z}}\right)^2
 \left({\nu\over\nu_{_R}}\right)^{5/2}
	\left({\epsilon_B\over 0.1}\right)^{-1/4} \left({E_{52}^{3/4} 
	t^{5/2} \over A_{\star} \Delta_{10}^{3/4} }\right) \ \rm{mJy},
\label{43}
\end{equation}
which increases rapidly with time as $t^{5/2}$. At the particular
time $t_{\rm cr}$ given in equation (\ref{24}) when the 
reverse shock crosses the inner edge of the coasting shell, the
black body flux is roughly
\begin{equation} 
F^{cr}_{\nu,\rm{bb}}\approx 640 \left({1+z\over 2}\right)^{11/2}
	\left({2-\sqrt{2}\over 1+z-\sqrt{1+z}}\right)^2 
\left({\nu\over\nu_{_R}}\right)^{5/2}
	\left({\epsilon_B\over 0.1}\right)^{-1/4} \left({E_{52}^{3/4} 
	\Delta_{10}^{7/4} \over A_{\star} }\right) \ \ \rm{mJy},
\label{44}
\end{equation}
which is roughly an order of magnitude above the (maximum) prompt optical 
flux at the same time, given by equation (\ref{25}), for typical 
parameters. Therefore,  it appears that synchrotron self absorption will
not affect our estimate of optical prompt emission from the reverse shock 
significantly, unless $\gamma_{3}$ and/or $\Delta_{10}$ are unusually 
small, and/or $A_\star$ and/or $E_{52}$ are exceptionally large. 
Finally, we note that the prompt emission estimated here may also be
lowered by inverse Compton scattering, which cools the radiating electrons 
in addition to synchrotron emission (Sari \& Piran 1999b)

\subsection{Comparison with the ISM Case}

Prompt emission from gamma ray bursts in a constant density
interstellar medium has been investigated in detail by Sari \&
Piran (1999a,b) and  M\'esz\'aros \& Rees (1999). For the ease
of comparison, we shall compute several key quantities of the reverse 
shock, using the same notations and under the same assumptions 
as the wind case. We limit our discussion to
the case of a relativistic reverse shock, which demands that 
the relative Lorentz factor between the shocked and unshocked shell 
materials 
\begin{equation}
\gamma_{34}=1.5\left({2\over 1+z}\right)^{1/4}{\gamma_3 n^{1/8}
	\Delta_{10}^{1/8}\ t^{1/4}\over E_{52}^{1/8} },
\label{ism1}
\end{equation}
be greater than unity. We assume that the shell $\gamma_3$ is 
large enough (typically of order unity) 
so that the reverse shock becomes relativistic a few seconds after 
the explosion. The quantity $n$ in the above equation is the
number density of the ambient medium in units of cm$^{-3}$. 

The Lorentz factor for the forward shock is much higher. We find
\begin{equation}
\gamma_{12}=320\left({1+z\over 2}\right)^{1/4}{E_{52}^{1/8}
	\over n^{1/8}\Delta_{10}^{1/8}\ t^{1/4} },
\label{ism2}
\end{equation}
which decreases gradually with time. To account for the 
deceleration, we have adopted a relation
between the observer's time $t$ and the time $T$ in the rest frame 
of the origin of $t=(1+z)T/(4\gamma_{12}^2)$ instead of equation 
(\ref{15}).  Assuming a purely hydrogenic
interstellar medium (with $X=1$), we obtain a typical frequency 
in the reverse shock of
\begin{equation}
\nu_{m3}=1.6\times 10^{14}\left({3p-6\over p-1}\right)^2\left({2\over
	1+z}\right)\left({\epsilon_{e3}\over 0.1}\right)^2
	\left({\epsilon_B\over 0.1}\right)^{1/2}\gamma_3^2
	n^{1/2}\ \ \ \rm{Hz},
\label{ism3}
\end{equation}
and a cooling frequency of
\begin{equation}
\nu_c=1.6 \times 10^{17} \left({\epsilon_B\over 0.1}\right)^{-3/2}
{\Delta_{10}^{1/2}\over E_{52}^{1/2} n \ t}\ \ \ \rm{Hz}.
\label{ism4}
\end{equation}
For typical parameters, we therefore have $\nu_{m3}<\nu_{_R}<\nu_c$,
and the flux density in the R-band is given approximately by
$$
F_{\nu_{_R},3}=\left({\nu_{m3}\over \nu_{_R}}\right)^{(p-1)/2}
	F_{\nu_m,3}=85\ \left({\nu_{m3}\over\nu_{_R}}\right)^{(2p-5)/4}
	\left({2\over 
	1+z}\right)^{1/4}\left({2-\sqrt{2}\over 1+z-\sqrt{1+z}}
	\right)^2
$$
\begin{equation}
	\times \left({3p-6\over p-1}\right)^{3/2} 
	\left({\epsilon_{e3}\over 0.1}\right)^{3/2}
	\left({\epsilon_B\over 0.1}\right)^{7/8}
	{\gamma_3^{1/2}n^{5/8}E_{52}^{5/4}\ t^{1/2}
	\over \Delta_{10}^{5/4} }\ \ \ \rm{mJy},
\label{ism5}
\end{equation}
where we have used equations (\ref{18})--(\ref{19}) to derive the
peak flux density, $F_{\nu_m,3}$,at the typical 
frequency.\footnote{We note here that the peak flux would be lower by
a factor of about 3 if the formalism of Wijers \& Galama 
(1999) is adopted.}  The factor involving the ratio
\begin{equation}
{\nu_{m3}\over\nu_{_R}}=0.36\left({2\over 
	1+z}\right)\left({3p-6\over p-1}\right)^2 
	\left({\epsilon_{e3}\over 0.1}\right)^2
	\left({\epsilon_B\over 0.1}\right)^{1/2}
	\gamma_3^2 n^{1/2} 
\label{ism6}
\end{equation}
in equation (\ref{ism5}) 
becomes unity for the canonical value of $p=2.5$. Note that the 
optical flux density increases with time as $t^{1/2}$, as in
the wind case. For typical parameters, it appears that the 
optical flux in the ISM case is higher than that in 
the wind case (by a factor of about $6$), and the emission peaks at  
a magnitude of about 10 when the reverse shock crosses the inner
edge of the freely coasting shell. Of course, other choices of 
parameters can change these numbers substantially. For example, 
if one adopts a shell Lorentz factor $\gamma_{\rm{sh}}$ of 300 instead 
of $10^3$, then the optical fluxes in the wind and ISM cases
would become nearly identical, both peaking at a magnitude of 
about 10.7.  Note that the prompt optical fluxes in the wind
and ISM cases have the same dependence on the energy $E_{52}$,
the shell width $\Delta_{10}$, and time $t$. They have opposite 
dependences on the shell Lorentz factor $\gamma_3$, the ambient
density, and the magnetic energy fraction $\epsilon_B$. 
Furthermore, whereas the optical flux in the ISM case increases 
fairly rapidly with the electron energy fraction ($\propto 
\epsilon_e^{3/2}$ if $p=2.5$), that in the wind case is
independent of $\epsilon_e$. 

The X-ray emission from the reverse shock of the wind case is
expected to be higher than that in the ISM case. For 
definitiveness, we consider the emission at 2 keV, with $\nu_{_X}
=4.8\times 10^{17}$Hz. In the wind case (with $X=0$), we have for 
typical parameters $\nu_c<\nu_{_X}<\nu_{m3}$, and an X-ray flux 
density of
\begin{equation}
	F_{\nu_{_X},3}=440 
	\left({2-\sqrt{2}\over 1+z-\sqrt{1+z} }\right)^2
	\left({\epsilon_B\over 0.1}\right)^{-1/4}
	{E_{52}^{5/4}t^{1/2} \over A_\star^{1/2} \gamma_{3}
	\Delta_{10}^{5/4} } \ \ \mu \rm{Jy},
\label{ism7}
\end{equation}
which increases with time as $t^{1/2}$. For typical parameters, 
it has a peak value of $1.4$ mJy when the reverse shock crosses 
the inner edge of the freely coasting shell. It is interesting 
to note that the peak value is comparable to the flux 
density observed in GRB 970228 and GRB 970508, two of the GRBs 
to be discussed in detail in \S~4, on the timescale of tens of 
seconds, which is roughly 10 keV cm$^{-2}$ s$^{-1}$ (or about 
3 mJy at 2 keV; see Figures~1 and 2 of Frontera et al. 1999). 
The prompt X-ray emission from the forward shock should be 
weaker, as in the  case of prompt optical emission. 
Therefore, the prompt X-ray
emission in the wind case does not appear to violate any 
observational constraints. Indeed, for some GRBs, it may 
be able to account for at least part of the early X-ray emission 
detected by the Wide Field Camera of the BeppoSAX. We shall 
explore this possibility elsewhere. 

In the ISM case, we have for typical parameters $\nu_{m3}<\nu_c
<\nu_{_X}$ instead, and an X-ray flux density of 
(assuming $p=2.5$)
\begin{equation}
	F_{\nu_{_X},3}=260
	\left({2\over 1+z}\right)^{1/4}
	\left({2-\sqrt{2}\over 1+z-\sqrt{1+z}}\right)^2
	\left({\epsilon_{e3}\over
 	0.1}\right)^{3/2}\left({\epsilon_B\over 0.1}\right)^{1/8}
	{\gamma_3^{1/2} n^{1/8}E_{52}\over \Delta_{10} }\ \ \mu \rm{Jy},
\label{ism8}
\end{equation}
which is independent of time and is a factor of about 5 lower than the 
peak flux in the wind case for typical parameters. The contrast in
prompt X-ray flux between the wind and ISM cases would 
be higher if a smaller shell Lorentz factor $\gamma_{\rm{sh}}$ is adopted, 
keeping other parameters at their typical values.

\section{OBSERVED SOURCES}

\subsection {GRB 970228}

A recent development is the finding of evidence for a supernova in GRB 970228
(Reichart 1999; Galama et al. 1999b); the late spectrum is especially convincing.
At first sight, this appears to be in conflict with the apparently interstellar
nature of the afterglow: Fruchter et al. (1999a) found time evolution with
$\alpha=-1.10\pm 0.05$ at optical wavelengths and the overall evolution
appears to be compatible with expansion in a constant density medium
(Wijers, Rees, \& M\'esz\'aros 1997).
However, if the late observations are attributed to a supernova, the
decline of the nonthermal optical afterglow steepens and $\alpha=-1.58\pm 0.28$
(Reichart 1999) or $\alpha=-1.73_{-0.12}^{+0.09}$ (Galama et al. 1999b).
With a plausible amount of extinction, Reichart (1999) finds
$\beta=-0.61\pm 0.32$ at optical wavelengths.
The optical-to-X-ray spectral index is  better constrained; Galama et al. (1999b)
find $\beta_{oX} = -0.780\pm 0.022$ at early times.
Cooling evolution ($\nu_c$ below optical wavelengths) is unlikely
because $\beta=-0.78$ would imply $p=1.56$.
With $\beta=-0.78$, an adiabatic blast wave in a constant density medium
implies $\alpha=-1.17$, but a blast wave in a wind implies $\alpha=-1.67$
(M\'esz\'aros et al. 1998; CL) in good agreement with the observed decline.
GRB 970228 can thus be plausibly added to the wind interaction category
with $p\approx 2.6$.

The evolution of the X-ray afterglow shows $\alpha=-1.33^{+0.11}_{-0.13}$,
$\beta = -0.96\pm 0.19$ up to day 4 at 2--10 keV (Costa et al. 1997)
and $\alpha=-1.50^{+0.35}_{-0.23}$ up to day 10 at 0.1--2.4 keV
(Frontera et al. 1998).
There is some evidence for a flatter rate of decline at X-ray frequencies
compared to the optical, which would suggest that $\nu_c$ is below
the X-ray regime.
Equation (\ref{a19}) shows that the X-ray emission should be in the cooling regime
during the observed time period for reasonable values of the parameters.
The spectrum may be somewhat flatter than $\beta=-0.78$ in the optical,
steepening to $\beta < -1$ in the X-ray regime.

GRB 970228 was monitored at radio wavelengths, but was not detected at limits
between 10 $\mu$Jy and 1 mJy over the first year
(Frail et al. 1998a).
In an interstellar interaction model, the peak flux, 
$F_{\nu,{\rm max}}$, should
remain constant and move to lower frequency with time.
Wijers et al. (1997) take $F_{\nu,{\rm max}}\approx 5$ mJy based on the 
X-ray flux
near the end of the initial $\gamma$-ray burst.
The radio emission clearly did not reach this peak flux.
Frail et al. (1998a) explain this result by claiming that 
$F_{\nu,{\rm max}}$ for the
afterglow could not be estimated from the initial burst and that 
$F_{\nu,{\rm max}}\approx 20-40$ $\mu$Jy.
This requires a substantial gap in the X-ray evolution between the initial
burst and the afterglow, and that the first optical observations have
occurred shortly after $\nu_m$ passed through optical wavelengths.
In a wind model, $F_{\nu,{\rm max}}\propto t^{-1/2}$ so there
is a drop in $F_{\nu,{\rm max}}$ from the time of the early X-ray observations
($t\approx 50$ s) to the time of the radio observations.
The radio limits can  be accomodated even with an initially high
X-ray flux in the afterglow.

\subsection{GRB 970508}

The optical afterglow of GRB 970508
 followed power law evolution from day 2 to day
$\gsim 100$ with $\alpha =1.141\pm 0.014$ (Galama et al. 1998a).
Galama et al. (1998b) compiled the radio to X-ray spectrum of the source
on day 12.1 (see also Wijers \& Galama 1999).
They estimated that the
 cooling frequency was at $\nu_c=1.6\times 10^{14}$ Hz or a wavelength
of 2 $\mu$m, i.e. a frequency just below optical frequencies.
Based on the day 12.1 spectrum, Galama et al. (1998b) estimate
that $p=2.2$, which yields $\alpha=-1.15$ and $\beta=-1.1$ for
$\nu >\nu_c$ in both the $s=0$ and $s=2$ cases.
The value of $p$ is supported by both the observed spectrum and
the rate of decline.
However, because of the similar optical and X-ray evolution for
ISM and wind interaction, radio observations
are crucial for distinguishing between the $s=0$ and $s=2$ cases.
Extensive radio data exist for GRB 970508, at 8.46, 4.86, and
1.43 GHz frequencies (Frail et al. 1997; 
Galama et al. 1998c; Waxman, Kulkarni, \& 
Frail 1998; Frail, Waxman, \& Kulkarni 1999b). 
Here, we fit these radio data using the 
thin shell model of Li \& Chevalier (1999) to determine the 
burst parameters. Our model treats synchrotron emission from a 
(trans-)relativistic blastwave propagating in an $r^{-2}$ 
medium. Synchrotron self-absorption and relativistic effects 
are included, but not cooling. As usual, we take a power-law 
distribution of electron energy, and adopt an energy index of 
$p=2.2$.

Since the observed radio frequencies are all below the cooling
frequency $\nu_c$ in the time interval of interest, fitting 
the radio data alone fixes only three out of four parameters, 
as discussed in Li \& Chevalier (1999). The additional
constraint comes from the observed flux in the R-band, which
is affected by cooling. Together, we find the best model
parameters to be: $\epsilon_e=0.2$, $\epsilon_{_B}
=0.1$, $E_{52}=0.3$, and $A_{\star}=0.3$ (which is in the 
expected range of a Wolf-Rayet star). The model fit is 
displayed in Figs. 2 and 3; the radio data, from
the VLA (Frail et al. 1999b), and the model are the same
in the two figures and are shown in two different ways for clarity.
The general evolution shown in the numerical model is along 
the lines of model D in \S~2.4, but it has gradual transitions
between the evolutionary phases.
This is due both to the smoothness of the transitions in the radio
spectrum and to the time lag effects which result in different parts
of the shell being observed at different evolutionary phases at
the same observer's time.

The initial rise of the model flux
 at 1.43 GHz has $F_{\nu}\propto t$, as expected
in the self-absorbed regime.
The model flux on day 6.2 does fall below the 1.43 GHz VLA observation
of $100\pm 30$ $\mu$Jy (Frail et al. 1999b).
However, monitoring at Westerbork at 1.4 GHz over the first 40 days
failed to detect the afterglow (Galama et al. 1998c).
The Westerbork observations have lower sensitivity than those 
at the VLA, but
a combined map of the 21 observations during this time period
yielded a flux of $33\pm 40$ $\mu$Jy.
Galama et al. (1998c) take this as evidence that the source is
well into the self-absorbed regime at early times, as it is in our model.

Both wind and interstellar relativistic spherical models predict
that $\nu_m\propto t^{-3/2}$ and the data over the range $1.4-86$ GHz
are in approximate accord with this expectation (Frail et al. 1999b).
However, the models differ in that $F_{\nu_m}$ decreases with time in
the wind case, but not in the interstellar case.
Fig. 3 shows that the observed decrease of $F_{\nu_m}$ is well reproduced
by the wind model.
Additional data are available  at 15 and 86 GHz, and Frail et al. (1999b)
find that $F_{\nu_m}\propto \nu_m^{0.40\pm 0.04}$, close to the
expectation of $F_{\nu_m}\propto \nu_m^{0.33}$ in the wind model.

After about day 100, the 4.86 and 8.46 GHz fluxes 
tend towards approximate power law declines (Fig. 2; Frail et al. 1999b).
Frail et al. (1999b) find that for $t> 90$ d, 
$\alpha_{8.46{\rm~GHz}}=-1.3\pm 0.1$ and $\alpha_{4.86{\rm~GHz}}=-1.1\pm 0.1$
and that the data at the two frequencies can be combined for $t> 110$ d
to yield $\alpha=-1.14\pm 0.06$ and $\beta=-0.50\pm 0.06$
(where $F_{\nu}\propto t^{\alpha}\nu^{\beta}$).
They suggest that ``$\beta$ undergoes an abrupt drop from positive to
negative values''  at $t\approx 100$ d.
Our model does not produce an abrupt drop in $\beta$, but we believe
that the data are consistent with a gradual change (see Fig. 3).
We attribute the fact that $\alpha_{4.86{\rm~GHz}}$ is deduced to be
larger than $\alpha_{8.46{\rm~GHz}}$ to the later transition from
flat evolution at the lower frequency.
Fig. 3 shows that the model light curve behavior does go to $F_{\nu}
\propto t^{-1.4}$ as expected for relativistic, spherical, adiabatic expansion
in a wind with $p=2.2$, but that it eventually becomes steeper.
The steepening is due to a transition to nonrelativistic expansion,
which is included in our numerical model (Li \& Chevalier 1999).
At $t=400$ d, equation (\ref{a5}) yields $\gamma =1.2$ for the low
frequency case.

The determination of $\nu_c$ was carried out as follows.
Approximately, the observed late time flux 
decline at 8.46 GHz follows
\begin{equation}
F_{8.46\rm~GHz}\approx 500 \left({t\over 100\ \rm{days}}\right)^{-1.4}\ \ 
	\mu\rm{Jy},
\label{c1}
\end{equation}
which implies a flux in the R-band (with $\nu_{_R}=4.5\times
10^{14}$Hz) of
\begin{equation}
F_{_R}=\left({8.46\times 10^9\over 4.5\times 10^{14}}\right)^{0.6}F_{8.46\rm~GHz}
\approx 0.73\left({t\over 100\ \rm{days}}\right)^{-1.4}\ \ 
	\mu\rm{Jy},
\label{c2}
\end{equation}
if the R-band frequency is below the cooling frequency $\nu_c$. 
The cooling frequency is estimated in equation (\ref{a9}) which, for
the inferred parameters, becomes
\begin{equation}
\nu_c\approx 1.4\times 10^{13} t_{\rm{day}}^{1/2}\ \ \rm{Hz}.
\label{c3}
\end{equation}
Setting $\nu_c=\nu_{_R}$, we obtain a cooling time $t_{_R}=
1000$ days for the R-band emission. After $t_{_R}$, $\nu_{_R}
>\nu_c$, and the R-band flux is given by equation (\ref{c2}). 
Before $t_{_R}$, we have $\nu_{_R}<\nu_c$, and the R-band 
flux is given instead by
\begin{equation}
F_{_R}= 0.73\left({t_{_R}\over 100\ \rm{days}}\right)^{-1.4}
\left({t\over t_{_R}}\right)^{-1.15}=82 t_{\rm{day}}^{-1.15}\ \ 
	\mu\rm{Jy}.
\label{c4}
\end{equation}
This expected flux is shown in panel d of Fig.~1. Our model fits 
are reasonable overall in all four frequencies, lending support 
to the wind interaction scenario for this GRB. 
X-ray observations are expected to be compatible with our model
because it is the same as an ISM model for $\nu>\nu_c$.
The evolution between 6 hr and 6 days can be approximately fitted
by $F_{\nu}\propto t^{-1.1\pm 0.1}$ (Piro et al. 1998) and the optical/X-ray
spectral index is compatible with $\beta_{oX}= -p/2=-1.1$
(Galama et al. 1998b).

Galama et al. (1998b) end their paper advocating an $s=0$ blast wave model
for GRB 970508 by describing three deficiencies of the model:
(1) $F_{\nu,{\rm max}}$ should be constant but is observed to decrease with time;
(2) $\nu_A$ is predicted to be time-independent, but is observed to
decrease with time, and the rise of the radio fluxes is slower than
expected; and (3) the decay after maximum at mm wavelengths is perhaps
somewhat faster than expected.
All of these problems are addressed by the wind model, which provides
approximate quantitative agreement with the observations.
Waxman et al. (1998) also noted the problems with a relativistic,
spherical, $s=0$ model, and suggested jet effects and a transition
to nonrelativistic expansion as possible solutions.
Frail et al. (1999b) developed these suggestions in more detail and
proposed three phases of evolution in a uniform density medium:
1) $2 < t < 25$ d, relativistic expansion in which the evolution appears
spherical (although it is a jet) because relativistic effects allow
only a part of the flow to be observed;
2) $25 < t < 100$ d,  jet spreading causes $F_{\nu_m}$ and the radio
   fluxes to drop below what would be obtained in an extension
   of 1); and
3) $100 < t < 450$ d, nonrelativistic, spherical expansion after
jet spreading is complete.
Frail et al. (1999b) show that the emission expected in phase 3
is consistent with the radio data, but the complete model has yet to
be calculated.
Relativistic time lag effects are likely to be significant at transition
times and a possible concern is that the optical light curve does not clearly
show evidence for the transition from phase 1 to phase 2.

Although the radio data appear to provide  support for the wind
model, the wind interpretation does give rise to possible problems
regarding synchrotron cooling.
In the wind model, $\nu_c$ increases with time; at optical wavelengths,
a transition is expected from cooling evolution to adiabatic evolution.
The opposite is true in $s=0$ models.
Galama et al. (1998b) cite evidence that $\nu_c$ evolves as expected
in an $s=0$ model.
They identify an observed optical spectral transition between 1.0 and 1.8 days
from $\beta=-0.54\pm 0.14$ to $\beta=-1.12\pm 0.04$ with the break
frequency $\nu_c$ passing through the $R_C$ band.
The wind model would predict cooling in the $R_C$ band at this early time
and no transition.
However, in the time before day 1.5, the observed light curve deviates 
strongly from standard afterglow
 models (e.g., Galama et al. 1998a; Fruchter et al. 1999b)
so that the early spectral index cannot be used as a constraint on models
for the later power law evolution.
Galama et al. (1998b) also cite moderately flat spectra between the
$K$ band (Chary et al. 1998) and $R_C$ band on days 4.3 and 7.3 as
supporting adiabatic evolution in the infrared at that time.
This is in conflict with the wind model, but we believe that the overall
weight of evidence supports the wind model.
In fact, Chary et al. (1998) find that their data are consistent
with a $t^{-1.2}$ decline and state that their data ``agree reasonably
well'' with a spectral index about --1, as found in the optical (and
expected in our model).
In the wind model, the optical spectrum should flatten and the light curve
should steepen at late times.

Our model fails to account for the optical evolution before day 2
(e.g., Galama et al. 1998a; Fruchter et al. 1999b).
At  early times, the blast wave is expected to be in the fast cooling regime
($\nu_m  >\nu_c$); this occurs at 500 s in the $s=0$ model (Galama et al. 1998b),
but at $\sim 1$ day in the wind model.
Although the sharp rise in the light curve may be related to this
transition, it is not possible to account for the rise in a straightforward
way.
In addition, Piro et al. (1999) have found possible evidence for 
redshifted iron line emission in the X-ray afterglow at an age
$\sim 1$ day.
In our model, the preshock density at that age is $\sim 1\times 10^{-23}$
g cm$^{-3}$, which is orders of magnitude smaller than the density
required for the line feature (Piro et al. 1999).
As noted by Piro et al. (1999), ordinary stellar mass loss cannot
account for the feature that they tentatively observe.

One expectation of the wind interaction model is that the event be
accompanied by a supernova.
However, Fruchter (1999) has found that a supernova like SN 1998bw added
to the power law decline of the nonthermal afterglow emission would give
an observable bump in the light curve at $t\sim 20-50$ days.
Such a bump is not seen (Fruchter et al. 1999b) and a supernova in
GRB 970508 would have to be about 1 magnitude fainter than
SN 1998bw (Fruchter 1999).
Considering possible differences in ejected mass, $^{56}$Ni mass, and
explosion energy, we believe that some variation in supernova
properties is plausible.

A remarkable property of our model is that it approximately reproduces
the radio light curves over the entire range of observations from ages
of 5 to 400 days.
There are  reasons
why deviations  from the model might be expected.
If the ejecta initially cover a small solid angle, steepening of
the light curve is expected as the blast wave slows  down and
there is eventually lateral expansion (\S~2.6).
Equation (\ref{a24}) shows that $\theta_o \gsim 1$ is 
required to avoid the steepening;
that is, the blast wave is spherical or nearly so.
Rhoads (1999) and 
Sari et al. (1999) reached a similar conclusion for GRB 970508 based
on an interstellar interaction model for the optical emission from the source.
In addition, the blast wave must remain within the steady stellar wind.
  From equation (\ref{a4}) and the blast wave parameters, the observed shock
has $R\approx 3\times 10^{18}$ cm.
The discussion in CL shows that the wind can plausibly extend to this
distance.

\subsection{GRB 990123}

The afterglow of GRB 990123 was briefly discussed by CL as a probable case
of ISM interaction.
The observations over the time period $0.01-1.5$ day (Kulkarni et al. 1999;
Galama et al. 1999a; Castro-Tirado et al. 1999) 
can be fitted by an ISM interaction
model with $p=2.5$ and optical wavelengths in the adiabatic regime
($\alpha=-1.12$; $\beta=-0.75$) and X-ray wavelengths in the
cooling regime ($\alpha=-1.38$; $\beta=-1.25$).
The observed steeper decline in the X-rays vs. the
$R$ band ($\alpha_X=-1.44\pm 0.07$ vs. $\alpha_R=-1.10\pm 0.03$, Kulkarni
et al. 1999) is expected for ISM interaction, but not for
 wind interaction.

A new feature observed in GRB 990123 was a 9th magnitude optical flash
overlapping with the GRB (Akerlof et al. 1999). Sari \& Piran (1999) 
and M\'esz\'aros \& Rees (1999) found that the properties of the 
optical emission over the first $\sim 15$ minutes could be modeled
by synchrotron emission from the reverse shock front resulting 
from interaction with a constant density, interstellar medium.
In \S~3 here,  we find it difficult to produce such a 
strong synchrotron, optical flash with standard parameters.   If 
we let $z=1.6$, $X=0$, and $\Delta_{10}=4$ (so that
the emission peaks around 50 s, as observed), and assume 
$A_\star=\gamma_{3}=1$ and $\epsilon_B=0.1$, then an explosion
energy of $5\times 10^{53}$ ergs is required to produce the 9th magnitude 
optical flash.  Such an energy is  nearly two orders of magnitude 
higher than the standard value, but it is not  a problem 
for GRB 990123, the brightest $\gamma$-ray burst with a well 
localized position. The estimated isotropic $\gamma$-ray 
energy alone for this source is
 $\sim 3\times 10^{54}$ ergs (Kulkarni et al. 
1999). The explosion energy could be  higher. 
Therefore, the magnitude of the optical flash does not provide
a clear discriminator between wind interaction and interstellar
interaction in this case. 

The observed temporal behavior of the optical flash of GRB 990123
is not compatible with the predictions of the wind interaction 
model. The flash is predicted to rise with time as $t^{1/2}$
(assuming that self-absorption is not significant,
which is true for the parameters listed in the 
preceding paragraph), 
whereas the observed rise is much steeper, close to $t^{3.7}$ 
evolution.\footnote{  We note here that the steep rise is not reproduced
in the simplest ISM case with a uniform freely coasting shell 
and a relativistic reverse shock either. To explain the steep
rise, one needs to take into account additional complications, such
as a nonuniform coasting shell or a transrelativistic reverse
shock. } 
In addition, the emission from the reverse shock is short-lived
 in a wind interaction model and is not
compatible with the observed $t^{-2}$ flux evolution over the
first 15 minutes (Akerlof et al. 1999,
Sari \& Piran 1999a; \S~3). 
The reason for the cut-off of the emission in the wind case is that
$\nu_c <\nu_m$, so that once the reverse shock front passes through
the shell the electrons rapidly cool and there is no more emission
from the reverse shock.
In the ISM case, $\nu_c >\nu_m$ and long lived emission with a power
law decline can occur (Sari \& Piran 1999a).

\subsection{GRB 990510}

The optical afterglow of GRB 990510 was well-observed in the first 4 days
after the burst (Stanek et al. 1999; Harrison et al. 1999).
The light curve could be well-fitted by an initial power law
$F_{\nu}\propto t^{\alpha_1}$ followed by a steepening to another
power law $F_{\nu}\propto t^{\alpha_2}$.
Although there was considerable overlap in the data that they used for
their analyses, Stanek et al. (1999) found $\alpha_1=-0.76\pm 0.01$
and $\alpha_2=-2.40\pm 0.02$ while Harrison et al. (1999) found
$\alpha_1=-0.82\pm 0.02$ and $\alpha_2=-2.18\pm 0.05$.
The initial flat evolution is strongly suggestive of adiabatic evolution
($\nu_m < \nu <\nu_c$) in a constant density medium ($F_{\nu}
\propto t^{-3(p-1)/4}$).
The initial decline rate measured by Harrison et al. (1999) is then
consistent with $p=2.1$ and with the later evolution being due to a
slowed jet with $F_{\nu}\propto t^{-p}$ (Sari et al. 1999).
The results of Stanek et al. (1999) suggest a somewhat  smaller value of $p$
from the early evolution and a somewhat larger value of $p$ from the later
evolution, but they are close to the expected evolution.
The value $p=2.1$ implies $F_{\nu}\propto \nu^{-0.55}$, which is consistent
with the spectrum $F_{\nu}\propto \nu^{-0.61\pm 0.12}$ measured from
BVRI photometry (Stanek et al. 1999).
The hypothesis of jet evolution is supported by radio data which are consistent
with the expected $F_{\nu}\propto t^{-1/3}$ evolution (Harrison et al. 1999).

Evolution in a constant density medium thus gives a consistent picture
for this afterglow.
Evolution in a wind cannot plausibly account for the early flat decline,
even taking into account the possibility of being in the strong cooling
regime (\S\S~2.2 and 2.3).

The steepening of the afterglow light curve to $\alpha_2=-2.2$
after 1.4 days should have facilitated the observation of a supernova
above the nonthermal emission.
Our model would predict the absence of such a supernova if it was an
interstellar interactor, but a test is difficult because of the high
redshift of the event, $z=1.619$ (Vreeswijk et al. 1999).
{\it HST} observations by Fruchter et al. (1999c) showed that on 17.9 June, 1999
(day 38.5) the afterglow flux was somewhat above the extrapolation of
the $t^{\alpha_2}$ decline, but the excess counts were a factor 7 below
what would be expected if a supernova like SN 1998bw were present.

\subsection{Discussion}

A summary of the results from this section is in Table 1.
In addition to the sources discussed here, we have added
GRB 980425 (Kulkarni et al. 1998; Li \& Chevalier 1999)
and GRBs 980326 and 980519 (CL).
The redshift
is given in column 2 for those for which it has been determined
(see the recent compilation of Ghisellini 1999 for references).
The next column gives the afterglow type, as determined by the 
afterglow evolution.
We claim that GRBs 970228 and 970508 are probable wind interactors,
although they have been widely interpreted in terms of interstellar
interaction in the literature.
The reasons for this are varied.
In the case of GRB 970228, the recognition of supernova emission in the
optical light curve yields a steeper nonthermal afterglow decline, in
line with expectations for wind interaction.
In the case of GRB 970508, the cooling frequency $\nu_c$ is somewhat
below optical wavelengths so that the high frequency 
evolution is the same for wind
and interstellar interaction.
However, the predicted evolution is different at radio wavelengths and
the radio data are compatible with wind interaction.
The radio data are crucial for typing this burst, and they
are suggestive of wind interaction for the afterglow of GRB 980519.

Although we believe that the radio data for GRB 970508 support a
wind interaction model, this is a controversial result.
Frail et al. (1999b) have proposed a uniform medium interaction
model, as discussed in \S~4.2.
In addition, the high frequency, steeply declining afterglows of
GRBs 980326 and 980519 can be interpreted in terms of jet evolution
(Sari et al. 1999).
If the jet model applies, either wind or interstellar interaction
could have taken place (see \S~2.6).
Detailed observations over a long time base are needed to clearly
discriminate between wind and interstellar interaction models.

In a paper that appeared after this paper was submitted, Livio \&
Waxman (1999)  addressed the problem of distinguishing wind and
interstellar interaction afterglows when jets are present.
They note that once the lateral jet expansion is complete, 
spherical, sub-relativistic expansion is expected (see \S~2.5).
This should lead to a flattening of the afterglow light curves.
If the steep decline of some afterglows is due to spherical wind
interaction and not to jet interaction,
flattening of the light curve would not occur  unless
the blast wave shock front reached the edge of the free wind
expansion region.

We have examined the data available on other sources, but have not
found more that can be  typed.
The afterglow of GRB 971214 was extensively observed at optical wavelengths,
but there is probably significant extinction in the host galaxy
and radio data are not available.
The afterglow of GRB 980329 was observed at radio wavelengths; the
8.3 and 4.9 GHz data of Taylor et al. (1998) over the first 30 days
appear to be in the self-absorbed regime.
The data appear to be better approximated by $F_{\nu}\propto t^{1/2}$
evolution (interstellar) as opposed to $F_{\nu}\propto t$, but there
are strong scintillation effects.
The higher frequency data are difficult to model.
Dai \& Lu (1998) suggested that GRB 970616 is a wind interactor based on
the steep decline with time indicated by two X-ray flux measurements;
we believe that more detailed observations are needed to make an identification.

Column 4 of Table 1 shows that the estimated electron energy spectral
index, $p$, varies over the set of bursts.
The probable error in these estimates, $\sim 0.1$, cannot account for
the range.
The high values of $p$ for GRBs 980326 and 980519 are reduced in jet
models for these sources (Sari et al. 1999), but a range in $p$ of
at least $\sim 2.1-2.5$ is difficult to avoid.
The lack of a universal value for $p$ calls into question the initial
assumption that the shock front accelerates electrons to a power
law above $\gamma_m$ that is constant with energy and with time.
The possibility of curvature in the spectrum and/or time evolution of $p$
should be examined as the data on GRB afterglows improve.

Column 5 indicates whether there is evidence for a supernova
like SN 1998bw in the emission from the source.
The case is clearest for GRB 980425 and its probable association with
SN 1998bw.
The data on GRB 970228 are compatible with a supernova very much
like SN 1998bw (Reichart 1999; Galama et al. 1999b).
This is also true for GRB 980326, although the observational case
is not as clear (Bloom et al. 1999b).
As discussed in \S~4.2, GRB 970508 does not show evidence for a supernova
like SN 1998bw.
Bloom (1999) has reported a similar situation for GRB 980519.
These facts   appear to contradict the expectations of our model.
However, in both cases a supernova
somewhat fainter than SN 1998bw can be accomodated.
Supernovae  associated with  GRBs can be expected to have a range of properties;
some variation in explosion energy, ejected mass, and mass of $^{56}$Ni are
all plausible.

The sixth column  lists whether a prompt, optical flash was
observed.
Such an event has been seen only in GRB 990123 and was attributed to
synchrotron emission from the reverse shock wave
(Sari \& Piran 1999a; M\'esz\'aros \& Rees 1999).
 In our picture, the optical synchrotron emission from the 
reverse shock wave of bursts with a wind type afterglow could 
have a magnitude comparable to, but would die off faster than, 
that of bursts with an interstellar 
type afterglow. 
The decay rate of the optical flash in GRB990123 is consistent
with an ISM type, not with a wind type.

The last column of Table 1 lists whether there is evidence for a jet
in the afterglow evolution.
The entry ``No'' indicates that the jet opening angle is 
$\theta_o \gsim 1$.
The case of GRB 970508 was discussed by Rhoads (1999) and Sari et al. (1999)
for ISM interaction and in \S~4.2 for wind interaction.
Li \& Chevalier (1999) found that semi-relativistic, spherically symmetric
models provided a good description of SN 1998bw/GRB 980425.
GRBs 990123 and 990510 show probable jet effects on a timescale
$\sim 1-2$ days (Kulkarni et al. 1999; Harrison et al. 1999; Sari et al. 1999).
The cases of GRBs 980326 and 980519 are controversial.
If the steep declines of their optical afterglows are interpreted as  jet
effects (Sari et al. 1999) then the jet effects occur at an earlier time
than for GRBs 990123 and 990510, but if the declines are interpreted as
wind interaction (CL) then the jet effects occur later if at all.

\section{TWO TYPES OF PROGENITORS}

In CL, we suggested that there are two types of GRB progenitors among
the well-observed sources: massive stars that have afterglows characteristic
of wind interaction and are likely to be accompanied by supernovae, and
compact binary mergers that have afterglows characteristic of
constant density, interstellar interaction and are not accompanied
by supernovae.
Our discussion here has  increased the wind group by
adding GRBs 970228 and 970508.
The pattern of association with a supernova is strengthened by the finding
of strong evidence for a supernova in GRB 970228.
 A new result here is that the prompt optical flash in a wind 
interaction GRB is expected to rise slowly and to disappear 
abruptly, in contrast to 
the observed temporal behavior of the optical flash of GRB 990123. The
contradiction 
points to an interstellar
interaction for this case, which supports the interpretation of the afterglow
light curve evolution by CL.

Another possible indicator of the GRB progenitor type is the location
in a galaxy (Paczy\'nski 1998).
For the cases in which imaging with  {\it HST} is available, the bursts
appear to be superposed on or near the optical disks of the host galaxies.
In the case of GRB 990123, the burst is at about 5.8 kpc 
(0.$^{\prime\prime}$67)
 from the center
of the host galaxy, near the edge of the optical disk (Bloom et al. 1999a).
In the case of GRB 990510, no host galaxy has been detected, even with
deep imaging (Israel et al. 1999; Fruchter et al. 1999c).
However, the production of a typical afterglow for the interstellar
interaction case probably requires that the burst take place in or near
a galaxy so that the surrounding density is sufficiently high.
For GRB 970508, which we have identified as a wind interactor, the
burst is within 0.$^{\prime\prime}$01, or 70 pc of the galaxy center
(Fruchter et al. 1999b).
Wind interaction implies a massive star progenitor and the source may
have been in a nuclear starburst region.
In the case of GRB 970228, the source is about 0.$^{\prime\prime}$5
from the galaxy center (Sahu et al. 1997), which is comparable to
the interstellar interaction case of GRB 990123.
Location does not appear to provide a clear discriminator for the two
types of bursts considered here.

We have argued that there are two types of burst progenitors, so the
GRBs associated with the two types might be expected to have distinct
properties.
We have examined the data on the GRBs themselves and have found no
clear distinction between those that give rise to wind or to ISM
type afterglows.
In both cases, the $\gamma$-ray light curves are complex and last
for 10's of seconds.
One implication is that the mechanism giving rise to the initial GRBs
does not depend on the external medium.
This supports the generally favored internal shock model for the GRBs
(e.g., Piran 1999).

Another implication is that, for both types, the source must be capable
of producing a long duration burst.
This is expected for a massive star origin because the collapse time of
the stellar core is on the order of 10's of seconds (MacFadyen \& Woosley 1999;
Fryer, Woosley, \& Hartmann 1999a).
Among the compact binary mergers that have been proposed (Fryer et al. 1999a
and references therein), the type that may satisfy the duration requirement
is the black hole - white dwarf merger (Fryer et al. 1999b).
In addition, this type of merger is expected to occur in or near the host
galaxy, again in accord with observations.
For the small number of sources we have discussed here, the rate of
massive star
explosions is not very different from the rate of compact star explosions.
However, the observed interstellar interactors are somewhat more distant and
luminous than the wind interactors
on average, and the occurrence of GRB 980425/SN 1998bw
indicates the presence of low energy wind interactors.
This would imply a higher rate of massive star explosions
per unit volume  than  of compact star explosions.
However, if jet effects are more important for interstellar interactors
(as suggested by Table 1),  the rate of interstellar interactors is increased,
so no clear conclusions are possible for the relative rates.

One of the unexpected results of this study is the possible evidence for
a relation between jet effects and the afterglow (and thus progenitor)
type.
Based on a small number of objects, the interstellar interactors show
evidence for significant jet effects, while the wind interactors with
massive star progenitors do not.
This suggests the intriguing possibility that the GRB engine generates
a collimated flow, but in the case of a massive star progenitor the flow
becomes uncollimated upon passing through the star.

\section{CONCLUSIONS}

The results of our work can be summarized as follows:

1.  The afterglow light curves
 resulting from the interaction of a GRB with a circumstellar
wind have properties that depend on the observation frequency.
At radio frequencies, there are two possibilities.
Above $\nu_{Am}$ (equation [\ref{a17}]), the light curve is of type D
(Fig. 1) and below $\nu_{Am}$, it is of type E.
Detailed integrations over a spherical blast wave show these segments,
but without sharp transitions.
At high frequencies (optical and X-ray), the expected light curve is
of type A (Fig. 1).
The evolution may make
a transition from cooling ($F_{\nu}\propto \nu^{-p/2} t^{-(3p-2)/4}$)
to adiabatic ($F_{\nu}\propto \nu^{-(p-1)/2} t^{-(3p-1)/4}$ ) evolution,
or may remain in one of these phases during the period of
observation depending on the parameters,
especially $\epsilon_B$.
The transition occurs earlier at lower frequency.
For typical parameters, the fast cooling phase (all electrons cool)
lasts for $\sim 1-2$ days, considerably longer than in the interstellar 
interaction case.
The light curve can be modified from the above expressions during this
early phase.

2. During wind interaction, the transition to 
nonrelativistic evolution occurs at $\sim 2$ yr
for typical parameters.
This is longer than in the interstellar interaction case because of
the low density in the outer parts of the wind.

3.    The optical synchrotron emission from the early reverse shock 
wave has a peak magnitude of $\sim 12$ for standard parameters.
It could be comparable to the prompt emission for the interstellar
interaction case. Radio emission is strongly self-absorbed during 
this phase. 

4.  The recognition of supernova emission in the optical radiation from
GRB 970228 (Reichart 1999; Galama et al. 1999b) 
implies that the decline of the nonthermal afterglow is
faster than previously thought and is compatible with 
adiabatic wind interaction.

5. The optical spectrum and decline over days $2-100$ of GRB 970508
suggest cooling evolution and are compatible with either interstellar
or wind interaction models.
However, the extensive radio data on this source strongly suggest
wind interaction.
In our model, the blast wave energy is $3\times 10^{51}$ ergs and
the wind mass loss rate is $3\times 10^{-6}\ml$ for a wind
velocity of $1,000\kms$.

6.  In the case of GRB 990123, the afterglow showed a steeper decline
with time in X-rays than in optical emission, as expected for interaction
with a constant density medium.
A prompt optical flash was observed from this object, which can be
plausibly attributed to synchrotron emission from the reverse shock 
front due to
interaction with the interstellar medium (Sari \& Piran 1999a,b).
We have found that interaction with a wind cannot produce 
the temporal behavior of the observed flash. 
The case for interstellar interaction appears to be strong.
We have also identified GRB 990510 as a likely interstellar
interactor based on its early time evolution.

7.  Among the 7 sources for which we have suggested identifications, 5 are
wind interactors and 2 are interstellar interactors, although two of
the wind cases can also be interpreted as jets in either a wind or the 
interstellar medium.
The interstellar interactors are somewhat more luminous and distant, but 
appear to have more significant jet effects, so no clear conclusions
can be drawn regarding the relative rates of the two types.
However, there does not appear to be a clear distinction in the initial
$\gamma$-ray burst properties of these different types.
This provides additional evidence for a burst model that does not depend on
environment, e.g., the internal shock model for bursts.

The main result of our paper is the presentation of evidence for
two types of GRB afterglows in different environments: a constant
density interstellar medium and the wind of a Wolf-Rayet star.
The types are not immediately distinguishable because at an
age of a few days, the preshock wind density is comparable to
an interstellar density.
 At an age of seconds, the preshock density is higher for the wind
case and we predict that the wind interactors have prompt
optical emission that die off faster than the interstellar 
interactors.
In addition, we expect the wind interactors to be accompanied by
a supernova event.
Future observations of GRBs and their afterglows should provide a
clear test of our model.

\acknowledgments
We are especially grateful to the referee, Re'em Sari, for
pointing out corrections and for constructive comments in his
detailed report.
Comments and correspondence from Eli Waxman and Dale Frail were
important for clarifying and correcting our model for GRB 970508,
although they do not agree with our model.
We also thank Andy Fruchter and Pawan Kumar for useful correspondence
and conversations.
Support for this work was provided in part by NASA grant NAG5-8232.

\clearpage

\noindent{Table 1. Properties of GRB Afterglows }

\vspace{1cm}

\begin{tabular}{ccccccc}
\hline
Burst &  Redshift &   Afterglow &  Spectral  & Supernova  & Prompt, Bright & Jet \\
   GRB  &    $z$   &     Type  & Index, $p$  & like SN 1998bw
     &  Optical Flash & \\
\hline
970228 &  0.695 &         Wind    & 2.6  &  Yes  & &  \\
970508 &  0.835 &          Wind    & 2.2  &  No  & & No  \\
980326 &       &          Wind    & 3.0  &  Yes  & &  \\
       &       &          Jet    & 2.2  &  Yes  & & Yes \\
980425 &  0.0085 &          Wind    & 2.5 &   Yes  & & No  \\
980519 &        &          Wind    &  3.0 &  No  &  & \\
       &       &          Jet    & 2.2  &  No  & & Yes \\
990123 &  1.60   &      ISM    &  2.5  &   &  Yes &  Yes \\
990510 &  1.619  &         ISM    & 2.1  &  No  &  & Yes \\
 \hline
\end{tabular}

\clearpage

\includegraphics[scale=0.70]{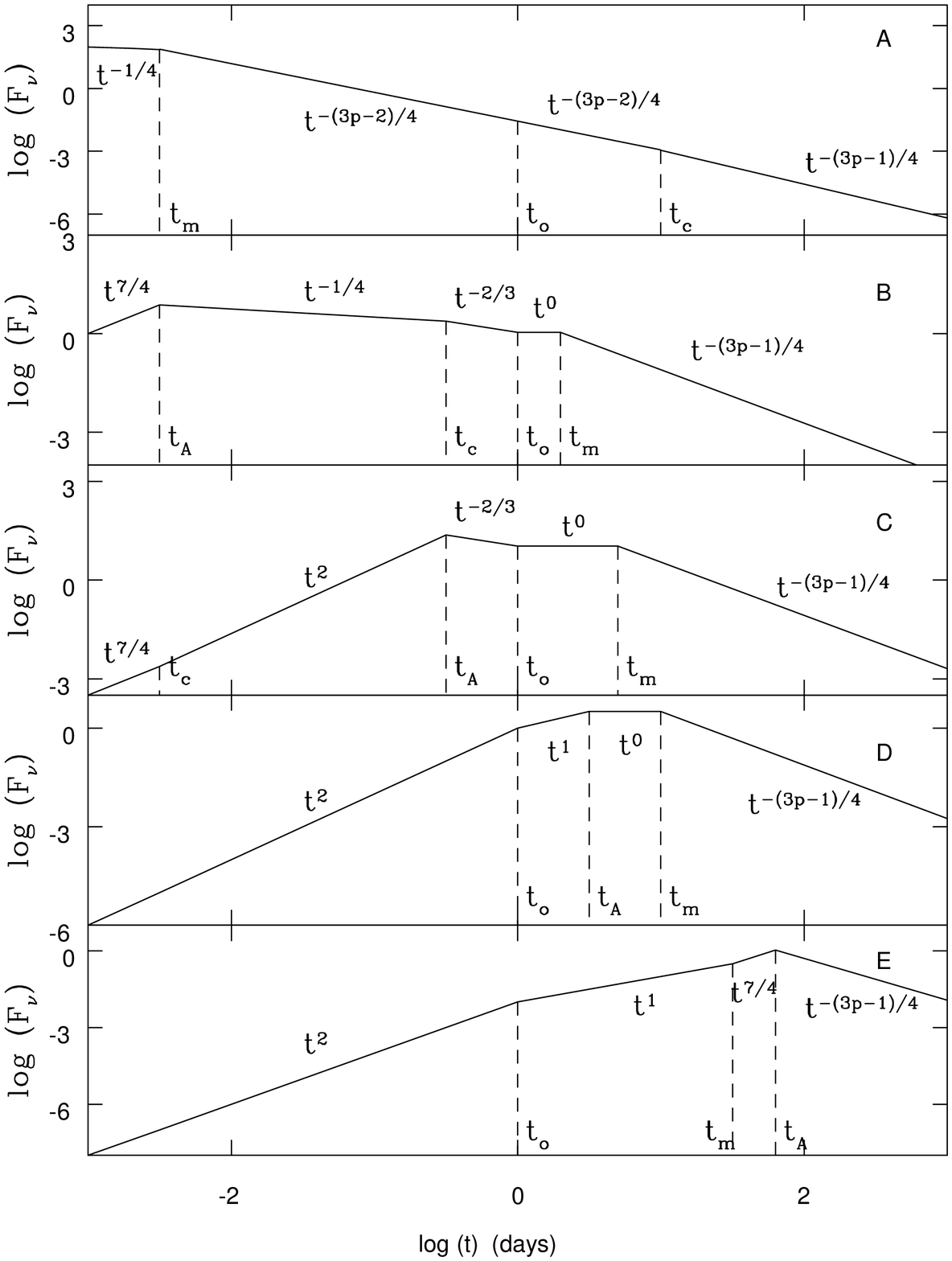}

\begin{figure}

\caption{Characteristic  light curves for wind interaction models in various
frequency ranges. 
The light curves are sorted from high frequency at the top (A) to
low frequency at the bottom (E).
X-ray and optical light curves are typically of type A and radio
of type D.}

\end{figure}

\clearpage

\includegraphics[scale=0.85]{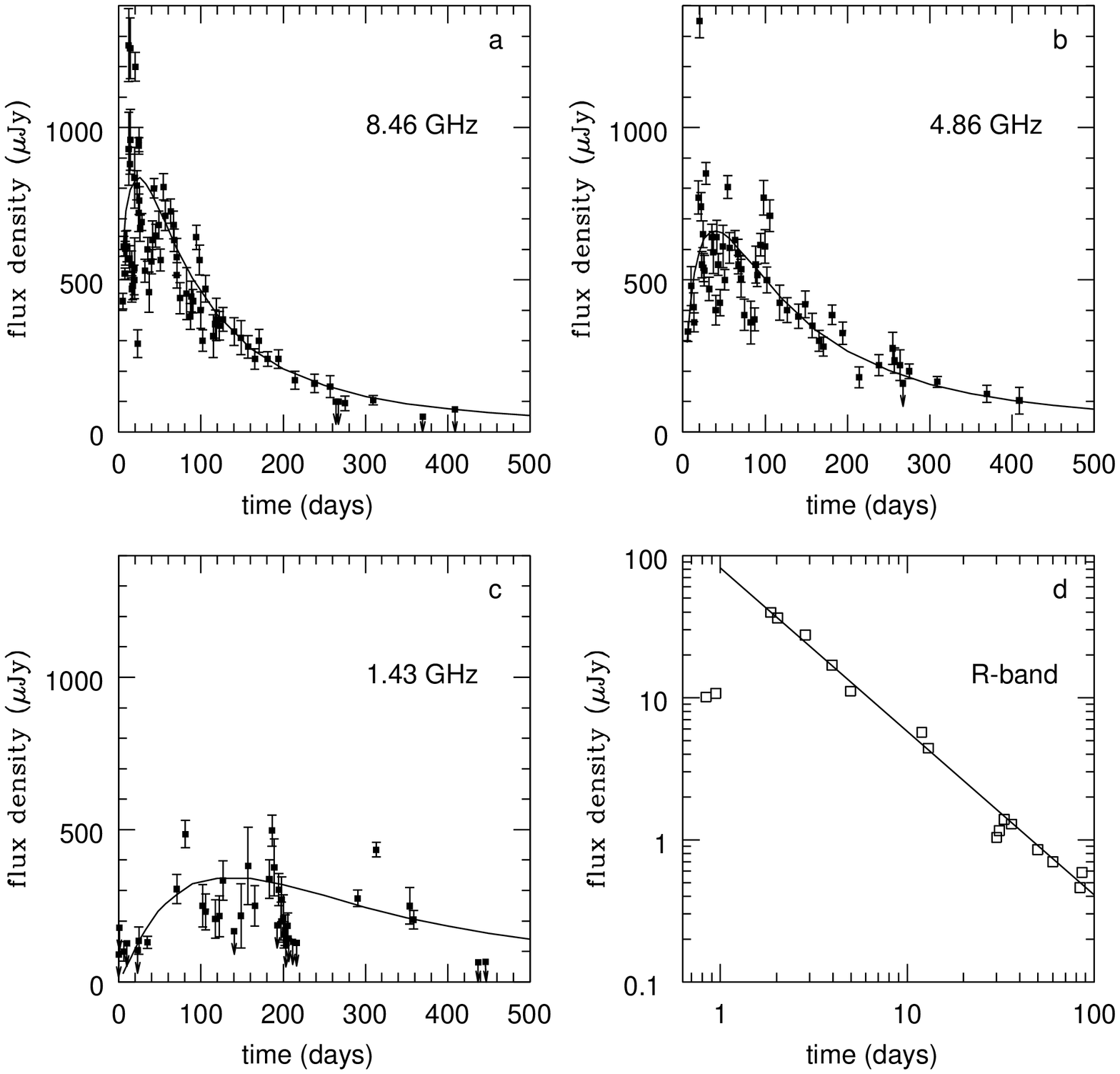}

\begin{figure}

\caption{ Wind interaction model for the afterglow of GRB 970508. 
Radio data  are taken from Frail et al. 
(1997; 1999b). R-band data are taken from 
Sokolov et al. (1998). The best model fit parameters are 
listed in the text.
}

\end{figure}

\clearpage

\includegraphics[scale=0.65]{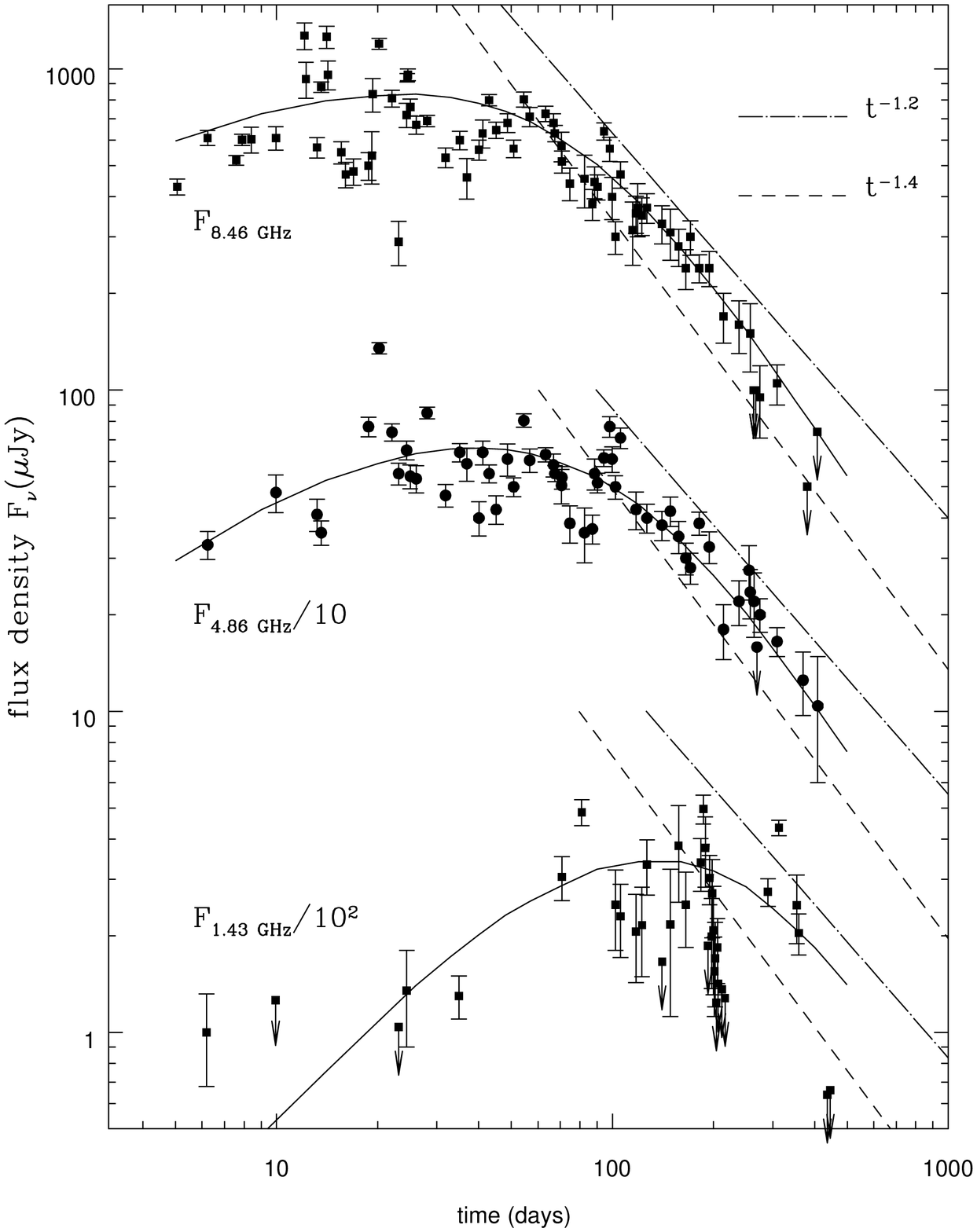}

\begin{figure}

\caption{ Log-log plot of the same observed data and model fits at the
radio frequencies as in Figure~2, showing the curvature in the 
power-law flux decay after about day 100. The data and fits for 
4.86 GHz and 1.43 GHz are artificially lowered by a factor of 10 
and $10^2$, respectively, for clarity.}

\end{figure}

\end{document}